%% file: Main.tex
\documentclass[]{aa}    % for a referee version
\usepackage{graphicx}
\usepackage{hyperref}
\usepackage{multirow}
\usepackage{float}	
\usepackage{textcomp}
\usepackage{subfigure}
\usepackage{gensymb}    % for \circ
\usepackage{natbib}
\usepackage{xcolor}
\usepackage{tikz}
\newcommand{\SgrA}{Sgr A$^{\ast}$}
\newcommand{\GBFD}{\textit{GBFD20}}

\begin{document}
	\title{Implications from 3D modeling of gamma-ray signatures in the Galactic Center Region}
	\author{J.\ Becker Tjus\inst{1}\inst{2} \and P.-S. Blomenkamp\inst{2}\inst{3} \and J.\ D\"orner\inst{1}\inst{2} \and Horst Fichtner\inst{1}\inst{2} \and Anna Franckowiak \inst{2}\inst{3}  \and  M.\ R.\ Hoerbe\inst{1}\inst{2}\inst{4} \and E.\ M.\ Zaninger \inst{1}\inst{2}}
	\institute{Theoretische Physik IV, Fakult\"at f\"ur Physik \& Astronomie,
		Ruhr-Universit\"at Bochum, 44780 Bochum Germany\\
		\email{julia.tjus@rub.de}
		\and 
Ruhr Astroparticle and Plasma Physics Center (RAPP Center), Ruhr-Universit\"at Bochum, 44780 Bochum, Germany	\and 
		Fakult\"at f\"ur Physik \& Astronomie, Astronomisches Institut, Ruhr-Universit\"at Bochum, 44780 Bochum, Germany	\and
University of Oxford, Oxford Astrophysics, Denys Wilkinson Building, Keble Road, Oxford, OX1 3RH, United Kingdom
}
	
	\abstract
	%state-of-the-art: 
	{The Galactic Center (GC) region has been studied in gamma rays in the past decades - the GC excess detected by Fermi is still not fully understood and the first detection of a \textit{PeVatron} by H.E.S.S.\ indicates the existence of sources that can accelerate cosmic rays up to a PeV or higher.}
	% aims:
{In this paper, we are investigating the origin of the \textit{PeVatron} emission detected by H.E.S.S.\ by, for the first time, simulating cosmic rays in the GC in a realistic three-dimensional gas and photon field distribution and large-scale magnetic field.}
% methods
{We solve the 3D transport equation with an anisotropic diffusion tensor using the approach of stochastic differential equations as implemented in the propagation software CRPropa 3.1 \citep{CRpropa2017}. We test five different source distributions for four different configurations of the diffusion tensor, i.e.\ with ratios of the perpendicular to parallel components $\epsilon=0.001,\,0.01,\,0.1,\,0.3$.}
% results
{We find that the two-dimensional distribution of gamma rays as measured by H.E.S.S.\ is best fit by a model that considers three cosmic-ray sources, i.e.\ a central source, the SNR G0.9+0.1 and the source HESS J1746-285. The fit indicates that propagation is dominated by parallel diffusion with $\epsilon=0.001$.  }
% conclusions
{We find that the 3D propagation in the 3D gas and B-field configurations taken from \cite{guenduez_bfield2020} can explain the general features of the data well. 
%Some small-scale features in the gamma-ray distribution  that do not exist in our simulation results point to the existence of additional molecular clouds not included in these simulations.
We predict that CTA should be able to identify emission from SgrB2 and the six dust ridge clouds that are included in our simulations and that should be detectable with a the expected resolution of CTA of $0.033^{\circ}$.}

	\maketitle
	\input{Contents.tex}
	\bibliographystyle{aa}
	\bibliography{literaturNew}
\end{document}

%% file: Contents.tex
\section{Introduction\label{intro:sec}}
The Galactic Center (GC) is one of the most interesting regions for the investigation of non-thermal processes in the Galaxy: The outflows as detected at gamma-ray \citep{FermiBubble2,FermiBubble3,Bubbles_Herold_2019,Bubbles_Fermi-LAT_2014}, microwave \citep{WmapHaze,WmapHazeDobler} and radio wavelengths \citep{RadioHalo}, are in need of proper modeling of the mass distribution and magnetic field structure in the GC. At GeV energies, the gamma-ray excess detected by the Large Area Telescope (LAT) on board of the \textit{Fermi} satellite is in need of explanation \citep{GC_Cholis_2021,GC_DiMauro_2021,GC_Pohl_2022} together with the very flat energy spectrum of $\sim E^{-2.3}$~\citep{FermiGalacticSpectrum} and at TeV-PeV energies, the the origin of the \textit{PeVatron} detected by H.E.S.S.\ needs to be understood~\citep{AbramowskiNature,HESS2018GC}. This detection indicates the existence of a PeV cosmic-ray source in the GC, thus for the first time providing a hint where in the Galaxy particles can be accelerated up to the knee of the cosmic-ray (CR) spectrum. In particular,
\cite{AbramowskiNature} report a spatially broad diffuse gamma-ray emission up to tens of TeV. Given that cosmic-ray interactions with gas lead to the production of gamma rays that receive approximately $10\%$ of the original cosmic-ray energy, it is clear that the underlying cosmic-ray source must have an energy spectrum that extends up to  the so-called cosmic-ray \textit{knee}, i.e.\ up to PeV energies. With a simplified approach of a one-dimensional diffusion modeling of the radial profile of the gamma rays and the gamma-ray spectrum, \cite{AbramowskiNature} come to the conclusion that there must be a central accelerator in the GC region with a cosmic-ray luminosity of $\sim 10^{37}-10^{38}$~erg/s.
An update of the analysis is presented in
\cite{HESS2018GC}, where a detailed morphological study was performed. The longitudinal and latitudinal distributions of the emission are presented and modeled with a many-component emission and gas model. It is shown that the H.E.S.S.\ data are best fit by a central source that produces the central peak. In addition, two point sources are identified in TeV gamma rays, i.e.\ at the location of the Supernova Remnant (SNR) G0.9+0.1 and at a location where a counterpart cannot be identified at this point, labeled HESS J1745--290.

It is discussed in \cite{HESS2018GC} that - due to the lack of a three-dimensional model of the ambient conditions in the GC Region - a three-dimensional simulation of the gamma-ray signatures was not possible so far. 
%until the publication of these results, no three-dimensional model of the ambient conditions in the Galactic Center Region was available, which is why a three-dimensional simulation of the gamma-ray signatures was not possible so far. 
A first three-dimensional modeling of the local gas distribution and the magnetic field in the Central Molecular Zone (CMZ) was presented in \cite{guenduez_bfield2020}. In that work, a magnetic field component for the central 200 pc has been derived from theoretical considerations based on existing knowledge on the diffuse gas component, molecular clouds (MCs) and non-thermal filaments (NTFs) in the CMZ. It could be shown that the general distribution of gamma rays is much better represented in this field configuration as compared to global magnetic field models that are typically used for propagation of cosmic rays through the Galaxy.

In this paper, we use the results of \cite{guenduez_bfield2020} - in the following referred to as \GBFD\ - in order to perform diffusive 3D cosmic-ray propagation in the CMZ. We include the three-dimensional gas distribution (diffuse gas and MCs) for interactions with the gas and propagate the particles using parallel and perpendicular diffusion in the three-dimensional magnetic field derived in \cite{guenduez_bfield2020}. We further use the distribution of stars as the source of a photon field, relevant for gamma-gamma interactions in the CMZ. Our propagation results are fitted to the H.E.S.S.\ data from \cite{HESS2018GC} and this way, we can test different hypotheses of the cosmic-ray source distribution and different models for the ratio of perpendicular to parallel diffusion coefficients. 

This paper is organized as follows: in Section \ref{ambient:sec}, we summarize the ambient conditions that we use for the propagation of cosmic rays in the CMZ. We present the propagation model in Section \ref{propagation:sec}, while results are shown and discussed in Section \ref{results:sec}. Finally, a summary and outlook is given in Section \ref{discussion:sec}.

%%%%%%%%%%%%%%%%%%%%%%%%%%%%%%%%%%%%%%%%%%%%%%%%%%%%%
\section{Three-dimensional modeling of the Galactic Center Region \label{ambient:sec}}
%%%%%%%%%%%%%%%%%%%%%%%%%%%%%%%%%%%%%%%%%%%%%%%%%%%%%
In order to model the three-dimensional transport and interaction of cosmic rays in the GC region, we use the mass distribution and magnetic field model presented in \cite{guenduez_bfield2020}. In this section, the most important features are summarized. In addition, we present a model for the photon field in the GC region, which is needed in order to properly consider gamma-gamma interactions.

%==================================================
\subsection{A model for the gas density in the GC}
%==================================================

The total density profile of the GC region is modeled to be composed of three individual components: 
\begin{enumerate}
    \item local Molecular Clouds (MCs) - the properties of the well-known clouds in the GC region are summarized in Table 1 in \cite{guenduez_bfield2020}.
    \item the gas structure of the innermost 10 pc of the GC region is taken into account as discussed in \cite{Inner10pc}. 
    \item an analytical description of the intercloud medium is given in \cite{MassGalacticCenter2}. As compared to that original publication, the normalization factor is scaled down in \cite{guenduez_bfield2020}, as the subtraction of the sub-structures that are considered separately (see item (1) and (2)) is necessary. This procedure results in the diffuse densities for H$_2$ and H as $n_{0,H_2}=91.1$ cm$^{-3}$ and $n_{0,H}=5.3$ cm$^{-3}$. We use these numbers for the diffuse gas component in this work as well.
    \end{enumerate}
    %==============================================
\subsection{A model for the three-dimensional large-scale magnetic field in the GC}
%==============================================
The total magnetic field is given by the superposition of three B-field components derived from
\begin{enumerate}
    \item the diffuse Inter Cloud Medium (ICM), $\vec{B}_{\text{ICM}}$
    \item eight known non-thermal filament (NTF) structures, $\vec{B}_{\text{NTF,i}}$ with $i=1,\ldots 8$;
    \item twelve local MC regions,  $\vec{B}_{\text{MC,j}}$ with $j=1,\ldots 12$.
    \end{enumerate}
The orientation of the large-scale magnetic field components in the ICM and the NTF regions is predominantly poloidal. The field is therefore modeled with an analytical, divergence-free and isotropic poloidal field model that is presented in \cite{X-ShapeModel} as \textit{model C}. Normalization, opening of the field lines and lengths scales are adapted individually for the eight NTFs and for the ICM as described in \cite{guenduez_bfield2020}. An average value of $\overline{B}_{\rm ICM}\approx$10 $\mu$G determines the field strength in the ICM \citep{FerriereMagneticField2009} and the typical parameters for the NTF regions can be taken from Table 1 in \cite{guenduez_bfield2020}.

The MC regions, on the other hand, are typically expected to have a horizontal field orientation. 
%Here, the ratio of the radial and azimuthal B-field components $B_r/B_\phi=\eta$ can be derived from the observational parameters for the gas density and the rotational velocity as described in \cite{CNRMagneticField}. 
The ratio $\eta = B_r / B_\phi$ between the radial and azimuthal field is set to $\eta = 0.5$ as suggested in \cite{guenduez_bfield2020}. The field structure is then derived by using Euler's $\alpha$ and $\beta$ potentials, which deliver a divergence-free expression of the magnetic field using $\vec{B}=\nabla\alpha\times \nabla\beta=(B_r,\,B_\phi,\,0)$.
The magnetic field strength in the MC region can be taken from Table 1 in \cite{guenduez_bfield2020}. 

The total field in the CMZ is then obtained by a superposition of the magnetic fields:
\begin{equation}
\vec{B}_{\text{tot}}=\vec{B}_{\text{ICM}}+\sum_{i=1}^{8}\vec{B}_{\text{NTF,i}}+\sum_{i=1}^{12}\vec{B}_{\text{MC,i}}\,.
\end{equation}

\begin{figure}[htb]
	\centering{
\includegraphics[width=\linewidth]{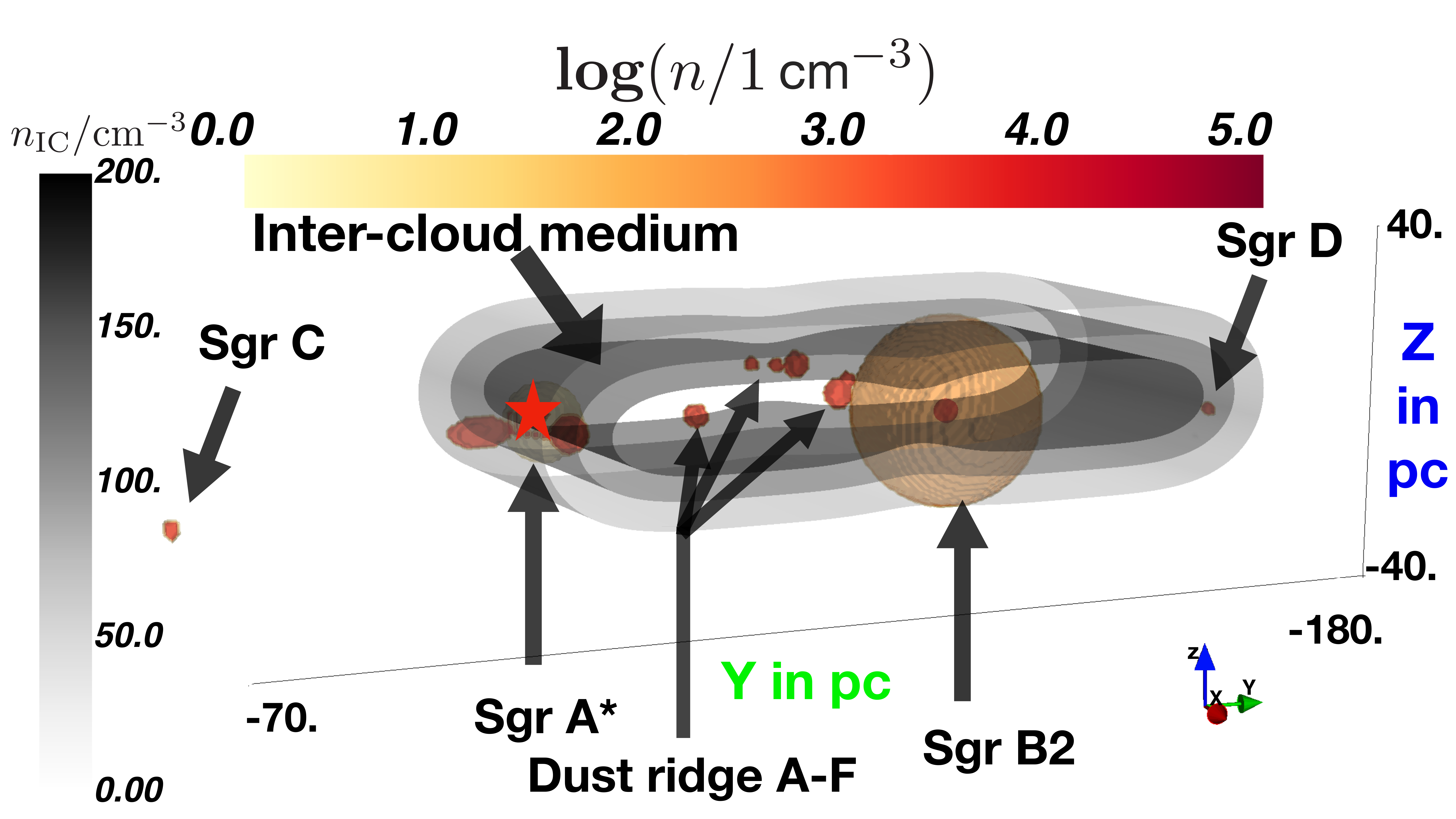}
}
\caption{Combined gas density of the CMZ region: the diffuse ICM is shown as gray contours and the contour represents a density level. \label{MassDistr1}}
\end{figure}

\begin{figure}
    \centering{
    \includegraphics[width=\linewidth]{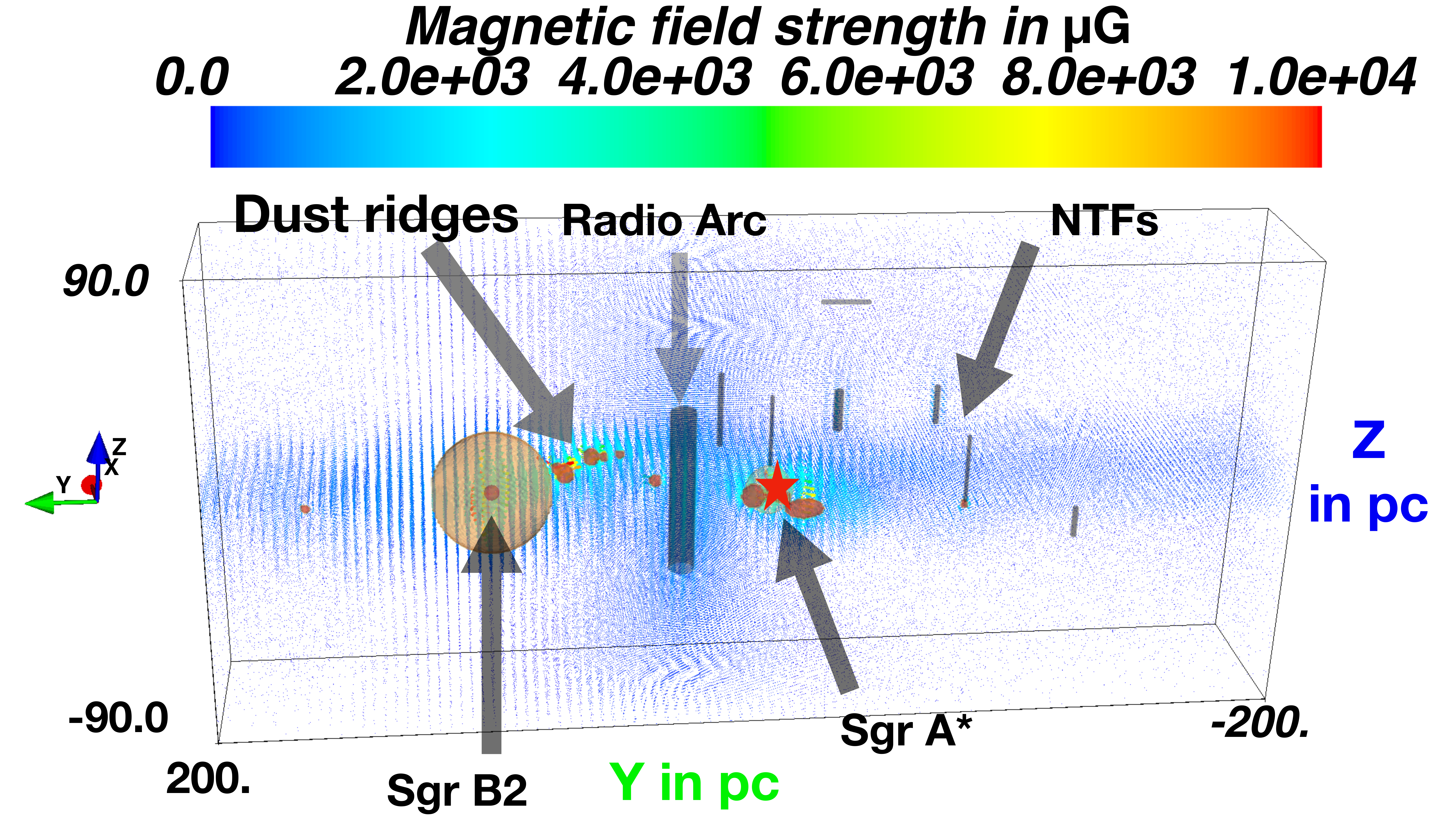}
    }
    \caption{The total magnetic field strength in the CMZ is given in $\mu$G. Additionally, NTFs are visualized by black cylinders.}
    \label{FieldTotalWithObjects}
\end{figure}

Figures \ref{MassDistr1} and \ref{FieldTotalWithObjects} display the 3D configuration and strength of the total mass distribution and magnetic field in the CMZ, respectively. These configurations are used in our simulations, using the gas distribution for modeling hadronic interactions and the large-scale magnetic field model for the three-dimensional propagation of the charged particles.


%==============================================
\subsection{A model for the background photon field in the GC}
%==============================================
\label{PhotonField}
The CMZ is home to more than 200 O-type stars \citep{StarFormationGC}. These types of stars are rare and bright, and their temperature exceeds 3$\cdot10^4$ K. For instance, 4 of the 90 brightest stars that are visible to the eye from Earth are O-type stars. They are prominent candidates for future violent supernova explosions due to their high mass. Thus, these types of stars represent energy reservoirs for CR acceleration in the CMZ. In total, three well-known star clusters in the GC exist: first the Central cluster, second the Quintuplet cluster and third the Arches cluster. Accordingly, the photon field is enhanced from FIR to UV wavelengths. Although in the CMZ the star and photon density is the highest in the Milky Way, star formation is suppressed, indicated by the lack of H$_2$O and methanol masers in the GC \citep{StarFormationGC}.
 \cite{StarFormationSupressionGC} show this suppression applies to the entire CMZ. \par
The background photon density has been discussed to play a crucial role for gamma-gamma interactions in \cite{Porter2018}. In addition, Inverse Compton scattering of high-energy electrons of the photon field can lead to the production of high-energy photons. And finally, high-energy photons undergo pair-production via their interaction with the ambient low-energy photons. In order to include these processes, this work makes use of a 3D photon distribution model as described below. Proton-gamma interactions are negligible considering the spatial extent and the energy range in the CMZ and are not included in this work.\par 

\begin{figure}[tb]
	\centering
	\includegraphics[width=1.0\linewidth]{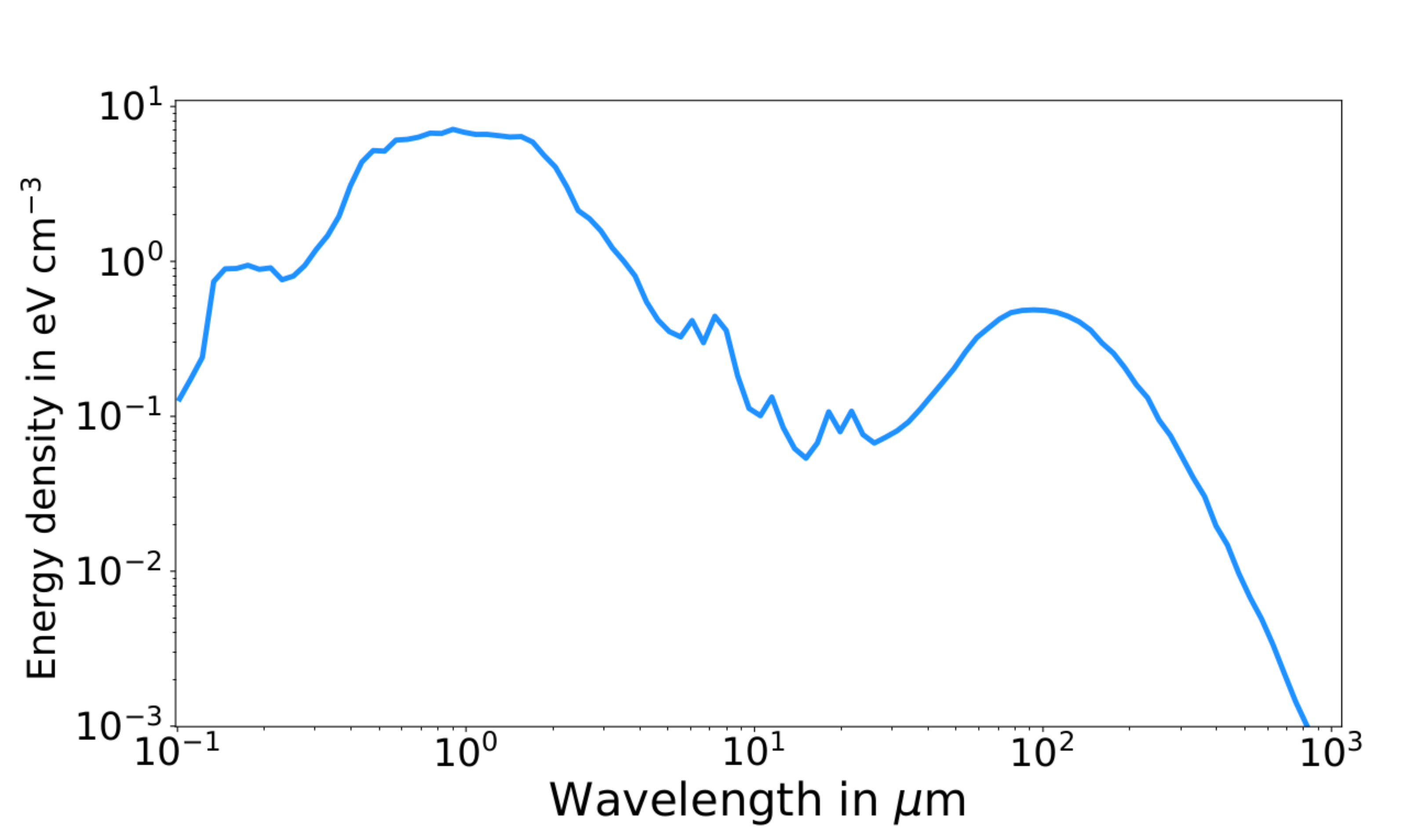}
	\caption[Spectral energy density of interstellar photons in the Galactic Center]{The spectral energy density of ISRF as a function of the wavelength for the model \cite{freudenreich1998} is shown here. Data are	taken from \cite{Porter2017}.
	}
	\label{fig:EnergyDensity2}
\end{figure}

\begin{figure}[tb]
	\centering{
	\includegraphics[width=1.0\linewidth]{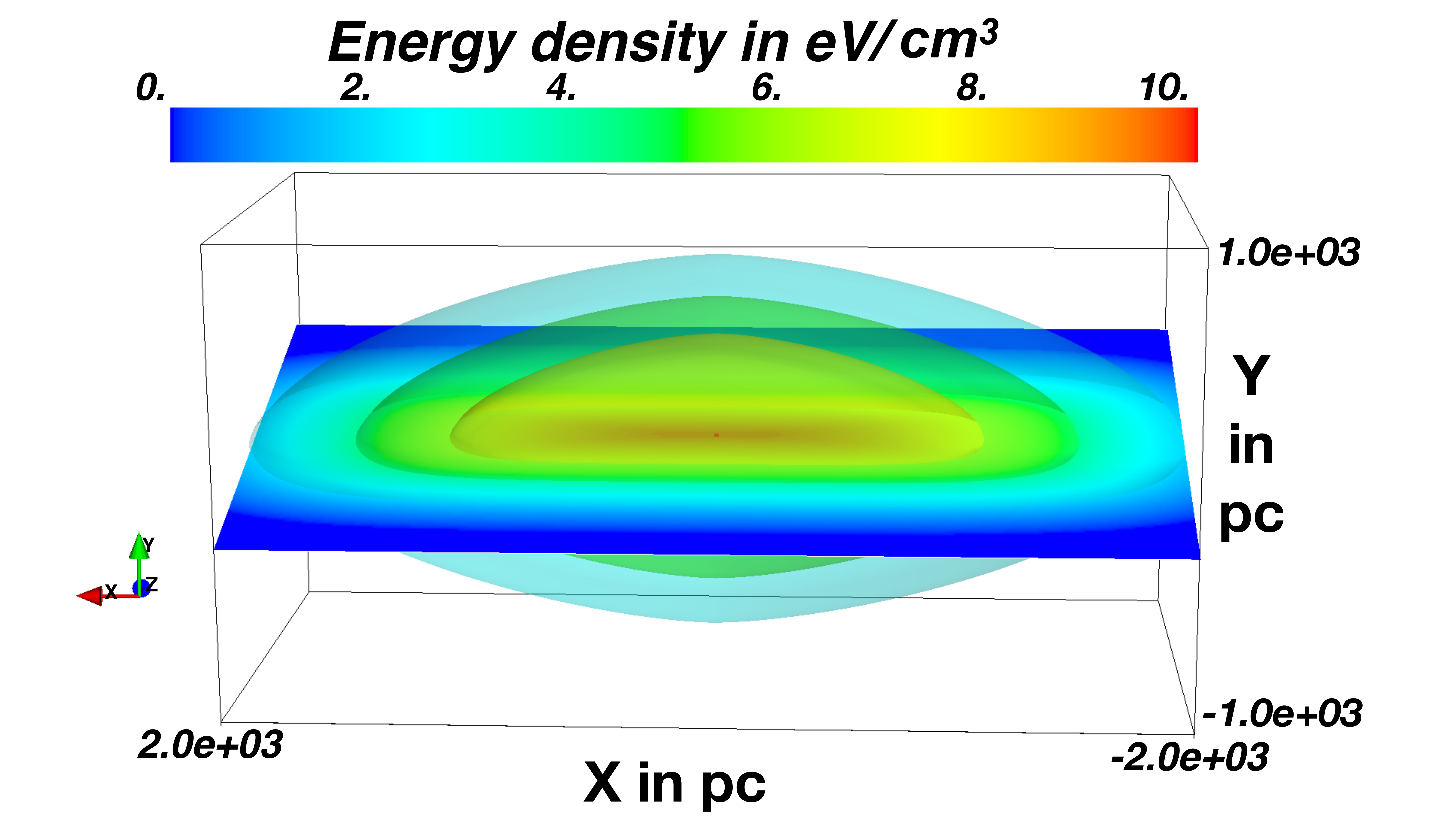}}
	\caption{The integrated energy density of ISRF as a function of spatial coordinates is shown for the  model presented in \cite{freudenreich1998}.
	\label{fig:EnergyDensity}}
\end{figure}

Sophisticated three-dimensional models of the interstellar radiation field (ISRF) in the Galaxy are presented by \cite{freudenreich1998} and \cite{robitaille2012}, both investigated in the context of cosmic-ray propagation in \cite{Porter2017,Porter2018}. While the model of \cite{robitaille2012} generally describes the \textit{global} Galactic emission best, it assumes local symmetry in the GC region. \cite{freudenreich1998} on the other hand considers a spatially asymmetric distribution of the photon field in the GC. This is why we apply the ISRF model developed in \cite{freudenreich1998} in our work.  
In the description of \cite{freudenreich1998}, the spatial description of the energy density is given by the function
\begin{align}
\label{eq:nPhoton}
n_{\mathrm{bulge}}(R, \phi, Z) = \ &n_{0} \operatorname{sech}^{2}\left(R_{s}\right) \nonumber \\
&\times \begin{cases} 1 &, R \leq R_\mathrm{end} \\
e^{-\left[\left(R_{s}-R_{\mathrm{end}}\right) / H_{\mathrm{end}}\right]} &, R > R_\mathrm{end}
\end{cases}
\end{align}
with $R_\mathrm{end} = 3.128$ pc, $H_\mathrm{end} = 0.461$ pc and the normalization factor  $n_{0}$ determined by the energy density and $R_{s}$ the bar radial coordinate
\begin{equation} 
 R_{s}=
 \left( 
 \left[ 
 \left(\frac{\left|X \right|}{ A_{X}}\right)^{C_{\perp}}
 +\left(\frac{\left|Y \right|}{ A_{Y}}\right)^{C_\perp}
 \right]^{C_{\|}} 
 +\left(\frac{\left|Z \right|}{  A_{Z}}\right)^{C_{\|}}
 \right) ^{1/C_{\|}} \, ,
\end{equation}
\begin{equation*}
\begin{split}
A_X=&1.6960 \ \mathrm{kpc,} \quad A_Y=0.6426 \ \mathrm{kpc,} \quad A_Z= 0.4425 \ \mathrm{kpc}\\
C_{\|}=&3.501\, , \quad  C_{\perp}=1.574 \, .
\end{split}
\end{equation*}
The photon energy density as a function of the wavelength is displayed in Figure \ref{fig:EnergyDensity2} and the resulting integrated energy spectrum as a function of spatial coordinates in Figure \ref{fig:EnergyDensity}.

Using this model, \cite{Porter2018} investigate the influence of gamma-ray attenuation due to electron-positron pair generation in the GC. They assert an attenuation of around $10\,\%$, which would lead to an increase in the true cut-off energy by a factor of $2.1$ as compared to what is measured with H.E.S.S. The authors assume an injection of photons from a single centralized source, which neglects the fact that the photons are more likely emitted diffusely through the entire CMZ. As the propagation length through the photon field in the GC region becomes shorter if they are produced with a certain distance from the center, it is likely that the attenuation is rather lower than the expectation presented in \cite{Porter2018}. We will test the influence of the photon distribution in this paper.

%%%%%%%%%%%%%%%%%%%%%%%%%%%%%%%%%%%%%%%%%%%%%%%%%%%%
\section{Simulation setup\label{propagation:sec}}
%%%%%%%%%%%%%%%%%%%%%%%%%%%%%%%%%%%%%%%%%%%%%%%%%%%%
For the propagation of cosmic rays in the GC region, we use the \textit{CRPropa} framework \citep{CRpropa2016}. While \textit{CRPropa} was originally written for the propagation of ultra-high-energy cosmic rays from extragalactic sources, it was extended by a solver of the transport equation in order to be able to treat lower-energy cosmic-ray propagation in galactic magnetic fields \citep{CRpropa2017}. A still existing limitation toward low energies comes from the fact that \textit{CRPropa}
 is relativistically formulated, i.e.\ the program makes use of the limit  $E=p\cdot c$ for the energetics of the particles and assumes that particle scattering of cosmic rays with a gas target is happening in the full forward direction. For this work, this means that particles with a Lorentz factor of $\gamma\ll 10^3$ cannot be taken into account and we therefore only consider cosmic-ray protons with $E_p>1$~TeV and cosmic-ray electrons with $E_e\gg$~MeV.
 
The transport equation implemented in \textit{CRPropa} is solved using the concept of Stochastic Differential Equations (SDEs) as described in detail in \cite{CRpropa2017}. The transport equation that is implemented in \textit{CRPropa} 3.1 is time-dependent and includes a diffusion term with a diffusion tensor $\hat{D}$ and an advection term with a velocity \vec{u},
\begin{equation}
    \frac{\partial n}{\partial t}=\nabla \cdot(\hat{D}\, \nabla n)-\vec{u} \cdot(\nabla n)+S(\vec{r}, p, t)\,.
\end{equation}
Here, $S(\vec{r},\,p,\,t)$ is the source term and $n$ is the local cosmic-ray differential intensity. The parametrization of the diffusion tensor is done by neglecting drifts and thus assuming a diagonal tensor in the frame of the local magnetic field line with a parallel and a perpendicular component $D_{\parallel}$. Their relation is parametrized via the constant factor $\epsilon:=D_{\perp}/D_{\parallel}$. In this paper, we will investigate the dependence of the result on the exact choice of $\epsilon$, as it is not clear from theory what the value should be. Diffusion is assumed to be in the quasi-linear limit for a Kolmogorov type wave vector spectrum, leading to $D_{\parallel}\propto E^{1/3}$. In particular at high turbulence level, there is evidence for a stronger energy dependence \citep{reichherzer2019, reichherzer2022}, but as the details about the turbulence level in the GC region are not clear, we use the 1/3-dependence. As we fit our results to the H.E.S.S.\ energy spectrum, the only thing that would change in our interpretation would be about the steepness of the source(s) that would need to be adjusted.

Within the module \textit{DiffusionSDE} in \textit{CRPropa}, this equation is solved by transforming the \textit{Fokker-Planck equation} to a set of parabolic SDEs. The SDE can then be treated numerically such that the next position of the pseudo-particle is determined. This way, interaction processes are considered at each step by calculating the probability for a specific interaction of the pseudo-particle with the ambient medium or field. These terms thus do not need to enter the transport equation and catastrophic as well as continuous loss terms can be considered properly this way. Due to the modular structure of \textit{CRPropa}, multiple propagation environments can be considered at the same time. In particular, the following variables are set:

\begin{enumerate}
	\item The numerical precision of the simulation is set to $P=10^{-3}$. While \cite{CRpropa2017} use values of $P=10^{-4}-10^{-5}$, it was shown in the same paper that one order lower in precision is enough to ensure the convergence of the 4th and 5th order of the Runge-Kutta algorithm that is used to solve the equations. 
	\item The minimum and maximum integration step length values are related to the minimum and maximum diffusion lengths in parallel direction -- defined by the upper/lower boundary of the wave vector spectrum in k-space, $l_{\min/\max}=2\pi/k_{\max/\min}$, respectively: In order to determine these, we use that in the SDE description, the root mean square of the deplacement is given by the Wiener process
	\begin{equation}
	    \langle x_{\nu}^2\rangle=D_{\nu\mu}\,d\omega^{\mu}\,,
	    \label{rms:equ}
	\end{equation}
	with $\nu=0,\,1,\,2,\,3$ and a summation over the index $\mu$ according to the commonly used sum convention. The term $d\omega^{\mu}$ is described by a four-dimensional Wiener process with Gaussian noise (see \citep{CRpropa2017} for details), $d\omega^{\mu}=\eta^{\mu}\sqrt{dt}$. Using this in Equ.\ (\ref{rms:equ}) for the parallel component, together with the resonance criterion for which scattering between particles with a gyroradius $r_g$ as a function of the rigidity $R=E/q$ happens for waves $k_{\rm res}\sim r_{g}^{-1}$, we get
\begin{eqnarray}
    l_{\mathrm{step,min}}=\sqrt{2\, D_{\parallel}(E=1 \ \mathrm{TV})}\cdot\sqrt{l_{\mathrm{min}}/c}\cdot\,  \eta_{\parallel}\,,\label{lstepmin:eq}\\
    l_{\mathrm{step,max}}=\sqrt{2\, D_{\parallel}(E=1 \ \mathrm{PV})}\cdot\sqrt{l_{\mathrm{max}}/c}\cdot \eta_{\parallel}\, .\label{lstepmax:eq}
\end{eqnarray}
Furthermore, $\eta_{\parallel}$ describes the randomness of the Wiener process according to the Gaussian noise, i.e.\ it denotes an independent normal-distributed random variable with zero mean and unity variance. 
The average absolute value for $\eta$ is $\langle|\eta|\rangle=0.8$. In order to receive values of $l_{\rm step,\min}$ and $l_{\rm step, \max}$ that are reasonably small, the $\eta_{\parallel}$ component used in Equ.\ (\ref{lstepmin:eq}) and (\ref{lstepmax:eq}) needs to be smaller than the average absolute value. We chose a conservative value of $\eta_{\parallel}=$0.2, which leads to a four times smaller step length and thus to more confident estimates.
\item Pseudo-particles are injected into the system and propagated and tracked according to their trajectory length. A stationary solution in the module SDE is achieved by simulating the discrete time-dependent solution and finally integrating over time \citep{CRpropa2017}. A detailed description of how to achieve the stationary from the time-dependent solution can be found in \cite{merten2018}.  In our simulation set-up such a solution is obtained for a maximum trajectory length of 5.1~kpc (see Figure \ref{fig:Stationary}).
\end{enumerate}
Within this work, the following changes and implementations have been considered, which have in parts been integrated in the new CRPropa 3.2 version \citep{CRPropa32}:
\begin{enumerate}
	\item 
	A new \textit{CRPropa} interaction module \textit{HadronicInteraction (HI)} integrates the generation of secondary electrons, neutrinos and gamma rays into the simulation. The interaction between CRs and the ambient medium via hadronic-pion production is taken into account. Here, the energy spectra of \cite{Kelner} and differential cross section of \cite{Kafexhiu2014} have been used. %\todo{cite Master thesis or github, Mario fuer das Improving!}

	\item 
	In order to use the \textit{HI} module adequately in the GC region, the mass distribution discussed in Section \ref{ambient:sec} is integrated into the framework of \textit{CRPropa} and the interaction module \textit{HI}. 
	\item
	The framework construction of electromagnetic interactions has been changed in order to consider customizable photon fields with any behavior as a function of energy, redshift, space, and time. Here, the space-time and redshift dependency are obtained by a separate scaling grid, which is multiplied with the energy-dependent photon density.
	\item The photon field energy density as a function of spatial coordinates and energy, as described in Section \ref{PhotonField}, is transcribed into the modified framework of \textit{CRPropa}. 
	\item
	A minimum Lorentz factor $\gamma_{\mathrm{min}}$ has been included into the \textit{Boundary} module.
\end{enumerate} 
We perform our simulations with the following final parameter settings:
\begin{enumerate}
    \item
    Each of the simulations is performed with $10^6$ particles on total, which follow a power-law distribution in the energy range $\Delta E=1~\mathrm{TeV}-1~\mathrm{PeV}$ with a spectral index of $\alpha=2.0$.

    \item
    In our final simulations presented in Section \ref{results:sec}, we further consider five source distribution scenarios presented below. For each of the five simulations, we distribute the particles correspondingly. 

    \item
    Simulations for four different values of the ratio of the perpendicular and parallel diffusion coefficient $\epsilon$ are produced, i.e.\ $\epsilon=0.001,\,0.01,\,0.1,\,0.3$.
\end{enumerate}

The magnetic field configuration is essential for the CR propagation in the GC region, as presented in Section \ref{ambient:sec}. The \GBFD\ model has also been integrated into the standard framework of \textit{CRPropa}, see \cite{CRPropa32}.

In the following subsection, we describe the exact settings in the simulation. For a better overview, Table \ref{tab:Input} summarizes all modules and the related input parameters as well as the outputs.
	\begin{table*}[ht]
		%\vspace*{-1.cm}
		\caption{A short overview of relevant CRPropa modules used in this work and the inputs and outputs concerned.}
		\centering
		\begin{tabular}{||c|c|c|c||} 
			\hline
			Module& Input &Input function -& Generated\\ 
			name &parameters&target field&particles\\
			\hline\hline
			\textit{Boundary}&$\gamma_{\mathrm{min}}=10^3$, $d_{\mathrm{max}}=5.1$ kpc&---&---\\
			\hline
			&$\left[\mathrm{Sgr A}^{\ast}\right]$, $\left[\mathrm{3sr}\right]$, $\left[\mathrm{uPSR}\right]$&&\\
			\textit{Source}& $\left[\mathrm{3sr+uPSR}\right]$, $\left[\mathrm{hom}\right]$&&$10^6$\\
			&$E=1$~TeV - $10^3$~TeV, $\alpha=2.0$&---&protons\\		
			\hline
			\textit{Diffusion} & $l_{\mathrm{min}}=1$~pc, $l_{\mathrm{max}}=5$~pc, $\eta=0.2$ &magnetic field&--- \\
			\textit{SDE}&$P=10^{-3}$, $\epsilon\in[0.3,0.1,0.01,0.001]$&\citep{guenduez_bfield2020} &\\
			\hline
			\textit{HI}&---&gas density&$\nu/\overline{\nu},\, e^{\pm},\, \gamma$\\
			\hline
			\textit{EMPP}&---&photon field& $e^{\pm}$\\
			&&GCB, CMB, CRB&\\
			\hline
			\textit{EMIC}&---& photon field & $\gamma$\\
			&&GCB, CMB, CRB&\\
			\hline\hline
		\end{tabular}		
		\label{tab:Input}
	\end{table*}
%==============================================
\subsection{Boundary conditions and basic assumptions}
\begin{figure}[htb]
	\centering
	\subfigure{\includegraphics[width=\linewidth]{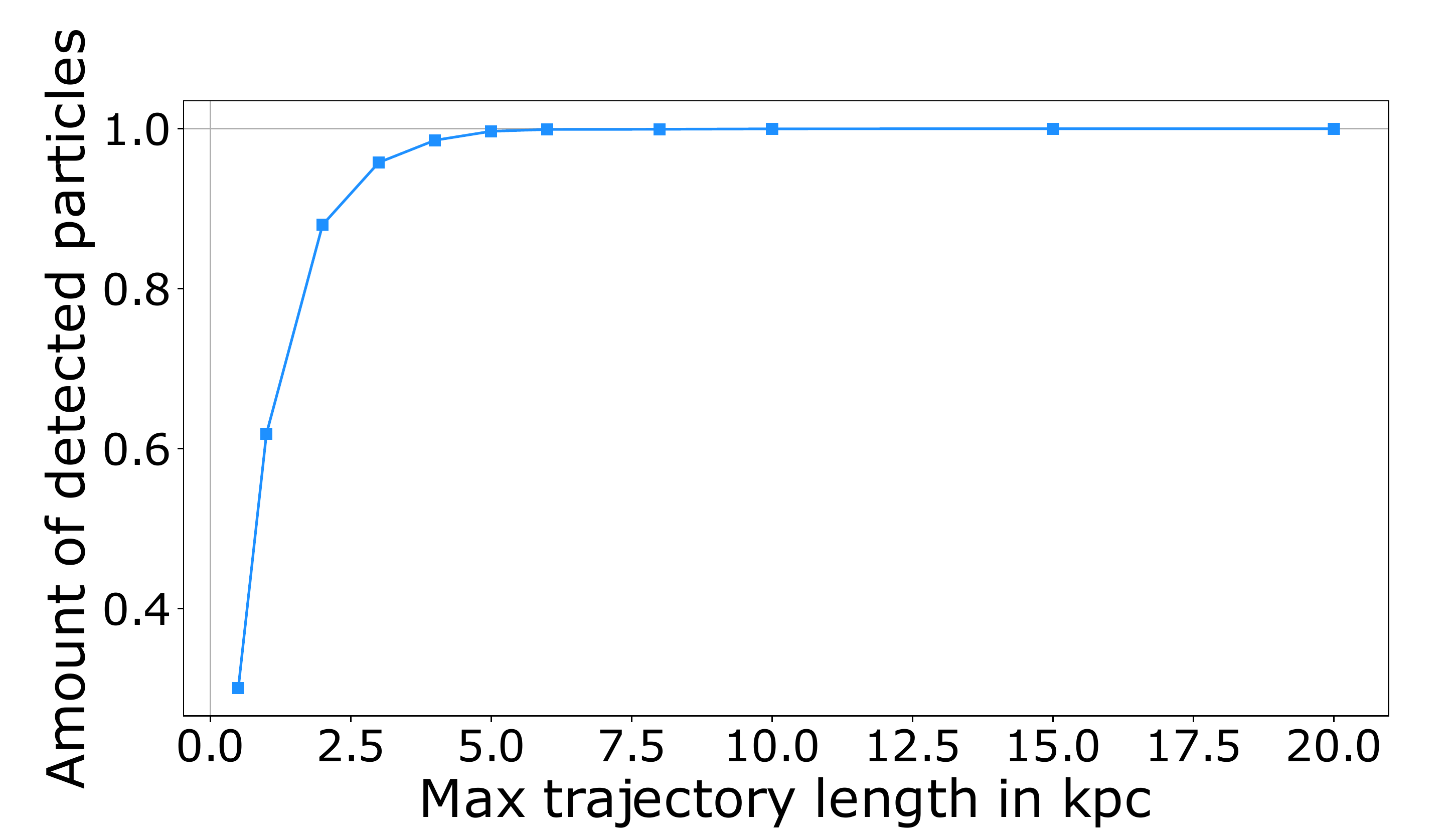}}
	\caption{The percentage of detected CRs after leaving a region of radius $R=200$~pc from a central source as a function of the maximum trajectory length is shown. In total, an injection of $10^6$ particles within the magnetic field configuration of \cite{guenduez_bfield2020} is considered.}
	\label{fig:Stationary}
\end{figure}

\begin{figure}[htb]
	\centering
	\includegraphics[width=\linewidth]{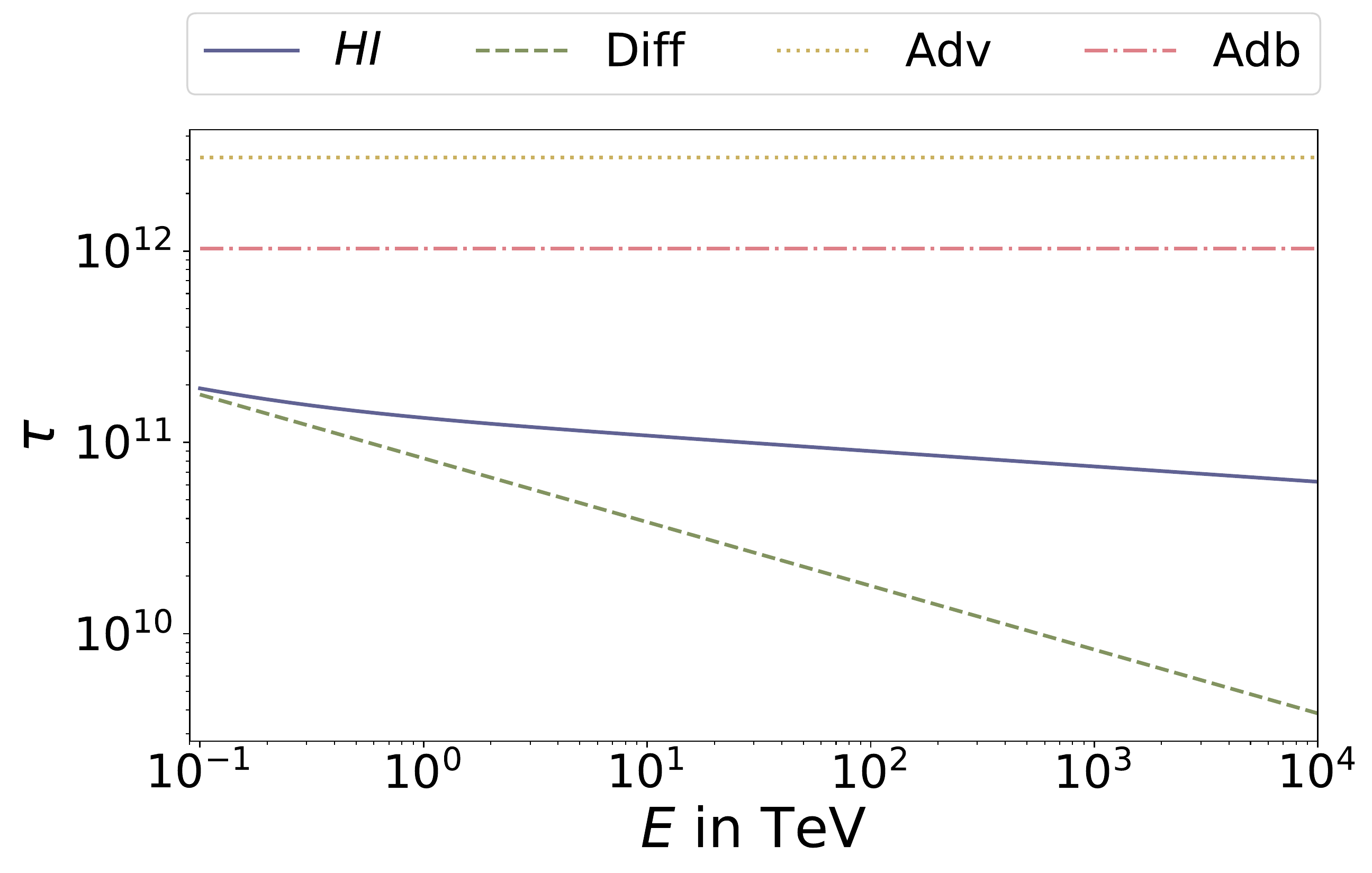}
	\caption[Loss timescale of different processes]{The loss timescales are shown for hadronic interactions (solid, blue line) as well as for diffusive (green, dashed line), advective (orange, dotted line) and adiabatic (red, dot-dashed line) losses. A region of $R=200$~pc and an average gas density of $10^4$~cm$^{-3}$ is used. The following abbreviations are used for labeling: HI=hadronic interactions, Diff=diffusion, Adv=advection and Adb=adiabatic loss. }
	\label{fig:Timescale}
\end{figure}

%==============================================
The simulation in this work is constrained by the following features:
\begin{enumerate}
	\item The minimum Lorentz factor is fixed to $\gamma_{\mathrm{min}}=1000$ so that the injected protons approximately move at the speed of light.
	\item The simulation volume is restricted to a box of $200 \times 400 \times 120$ pc. All particles leaving the simulation volume are lost.
	\item
	The maximum trajectory length $d_{\mathrm{max}}$ of injected CRs is related to the propagation time as $T_{\mathrm{max}}=d_{\mathrm{max}}/c$. A stationary solution is necessary in order to reproduce the observed diffuse gamma-ray emission \citep{HESS2018GC}. It is therefore useful to consider the CR propagation time required to leave the region of interest. Figure \ref{fig:Timescale} shows the different timescales for the propagation. The diffusion timescale (green, dashed line) at relativistic energies 
	is given as $\tau_{\mathrm{diff}}\simeq 3.3\cdot10^{12}\cdot (E/4\ \mathrm{GeV})^{-1/3}$~s  for traversing a region with radius $R=200$~pc as done in \citep{Guenduez18,GuenduezICRC2019} and dominates the other timescales of advection (orange, dotted line), adiabatic losses (red, dot-dashed line) and hadronic interactions (blue, solid line).	The maximum trajectory length for CR protons considering $\gamma_{\mathrm{min}}=1000$, i.e. $E\simeq 1$~TeV, yields
	\begin{equation}
	    d_{\mathrm{max}}=\tau_{\mathrm{diff}}\cdot c=5.1 \ \mathrm{kpc}.
	\end{equation}
    In fact, with these settings, almost all (more than 99$\,\%$) of the CRs injected from a centralized source leave the GC region. The amount of detected CRs after leaving the region of interest as a function of the maximum trajectory length is presented in Fig.\ \ref{fig:Stationary}.
	\item
	The source distribution $S$ is specified for the GC region as discussed in Section \ref{SourceDistribution}.

\end{enumerate}


%========================================================
\subsection{Optimization of $l_{\rm step, min}$ and $l_{\rm step, max}$}
\begin{figure*}[!ht]
\centering
\subfigure{\includegraphics[width=0.45\linewidth]{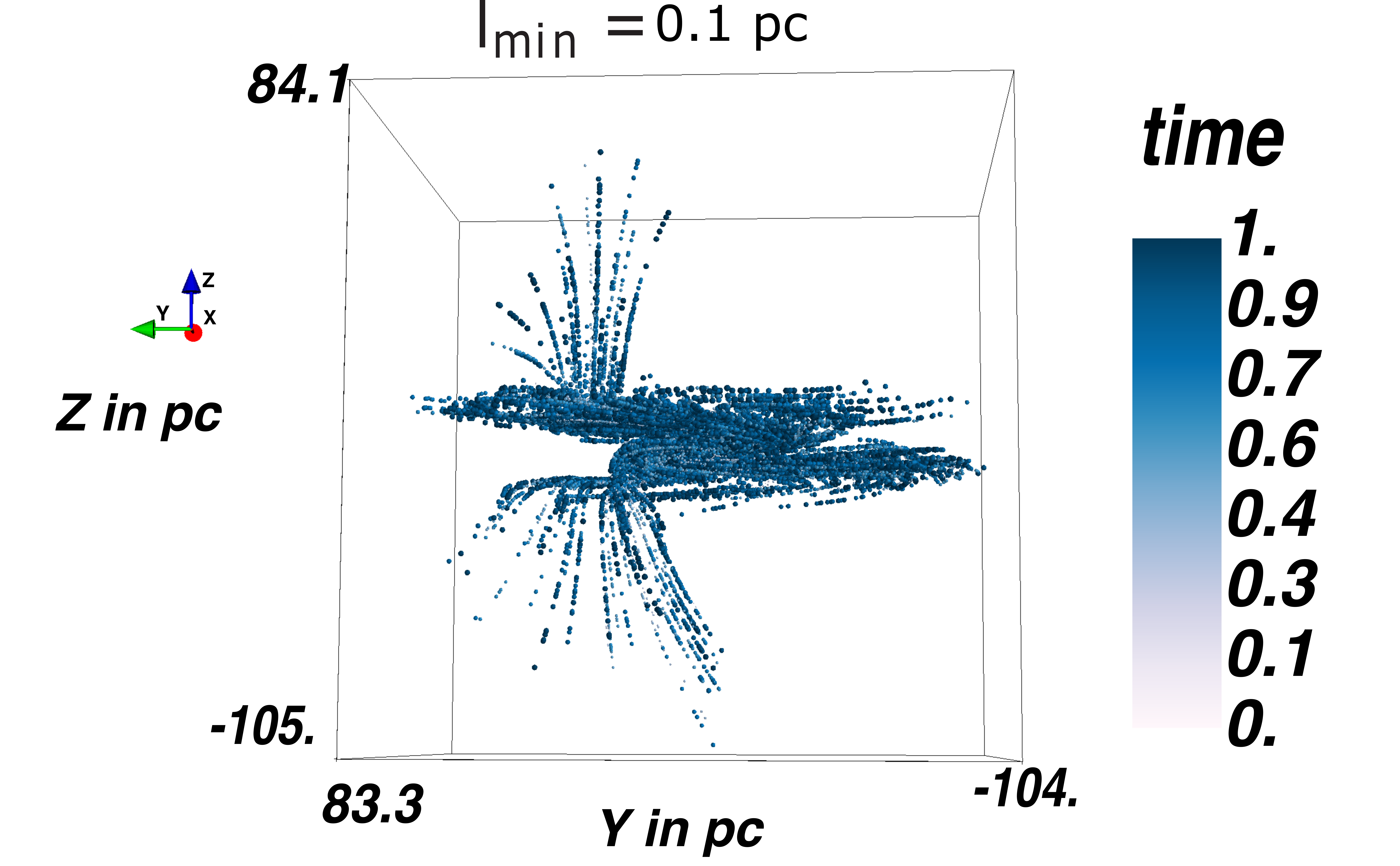}}
\subfigure{\includegraphics[width=0.45\linewidth]{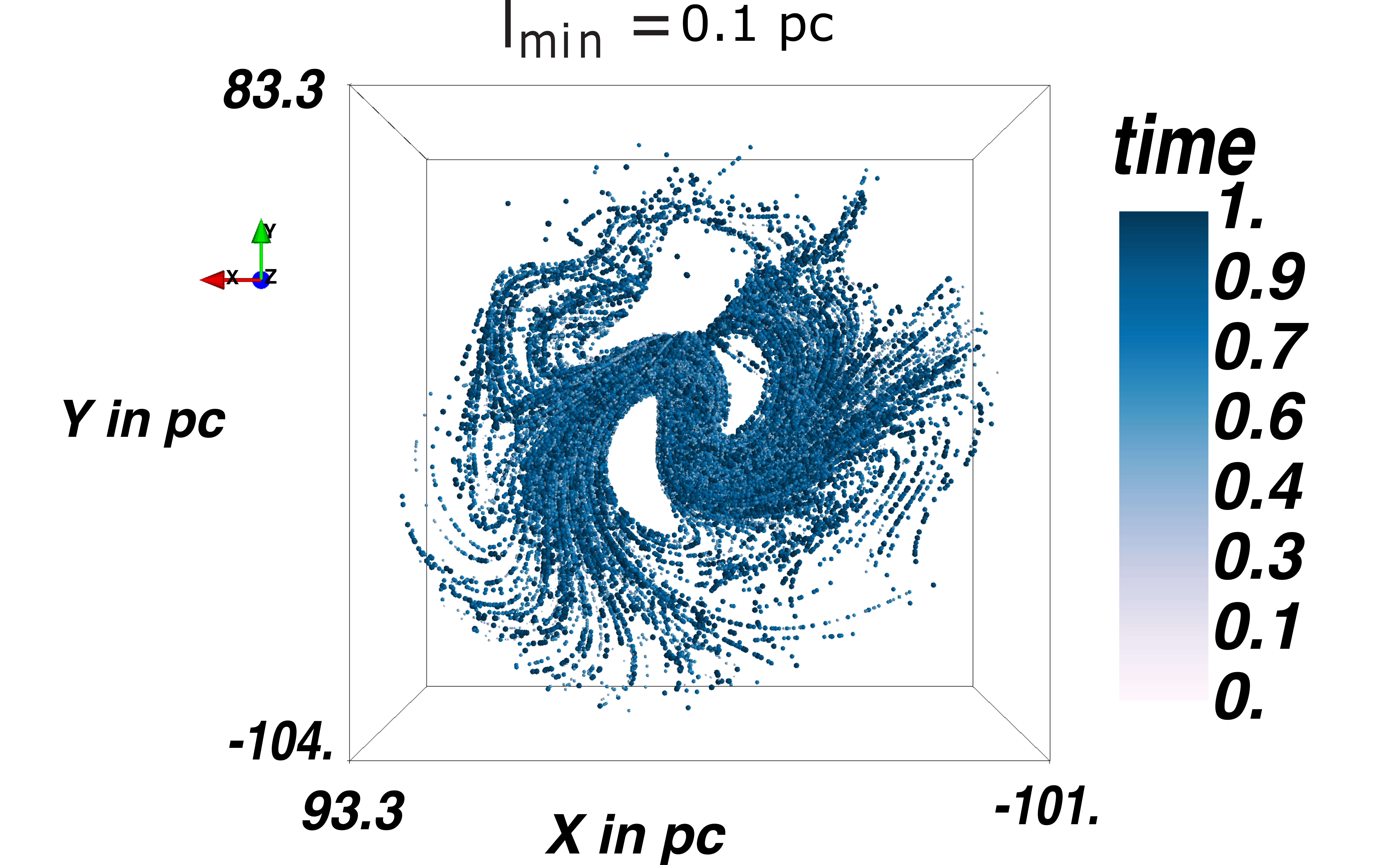}}
\subfigure{\includegraphics[width=0.45\linewidth]{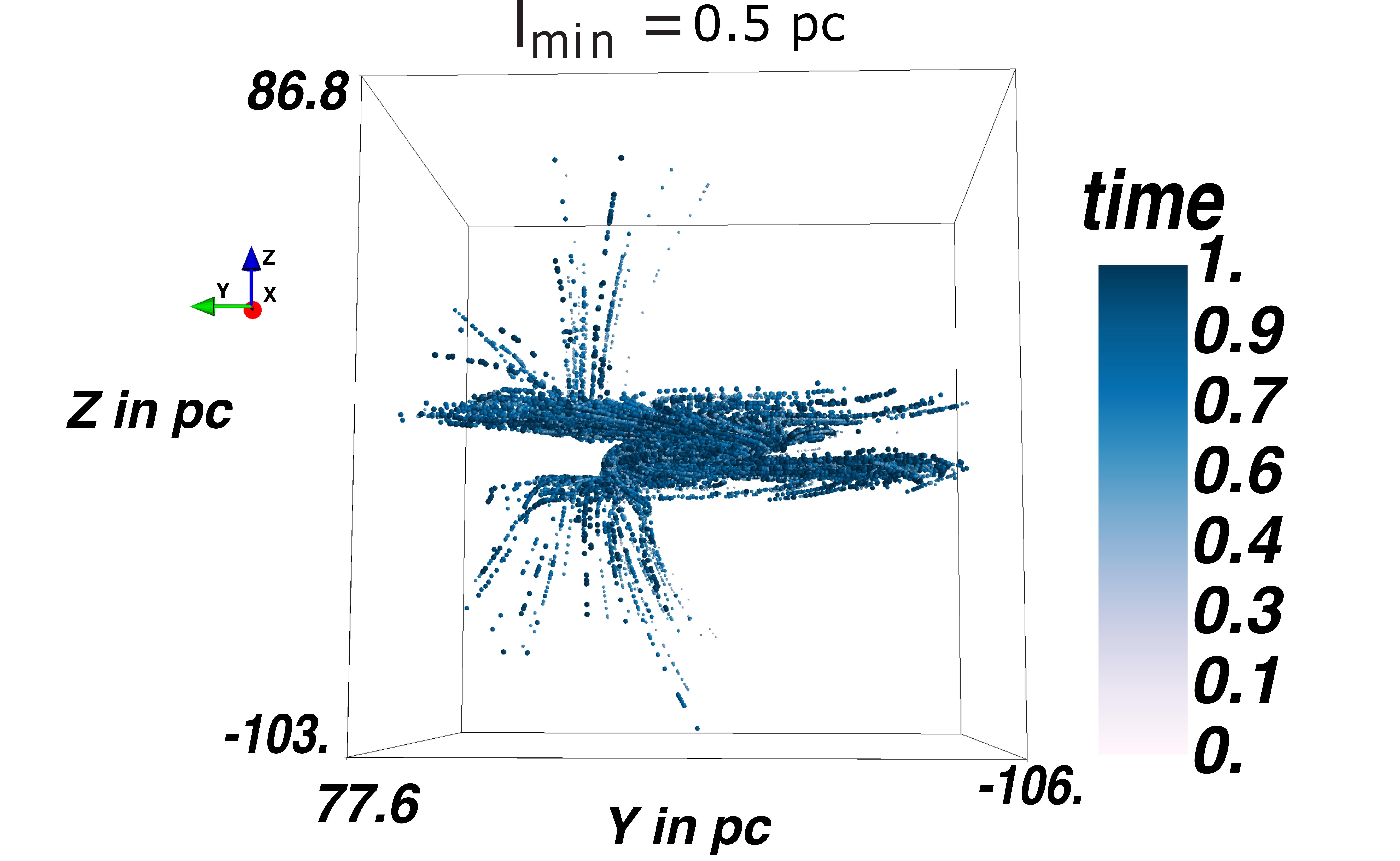}}
\subfigure{\includegraphics[width=0.45\linewidth]{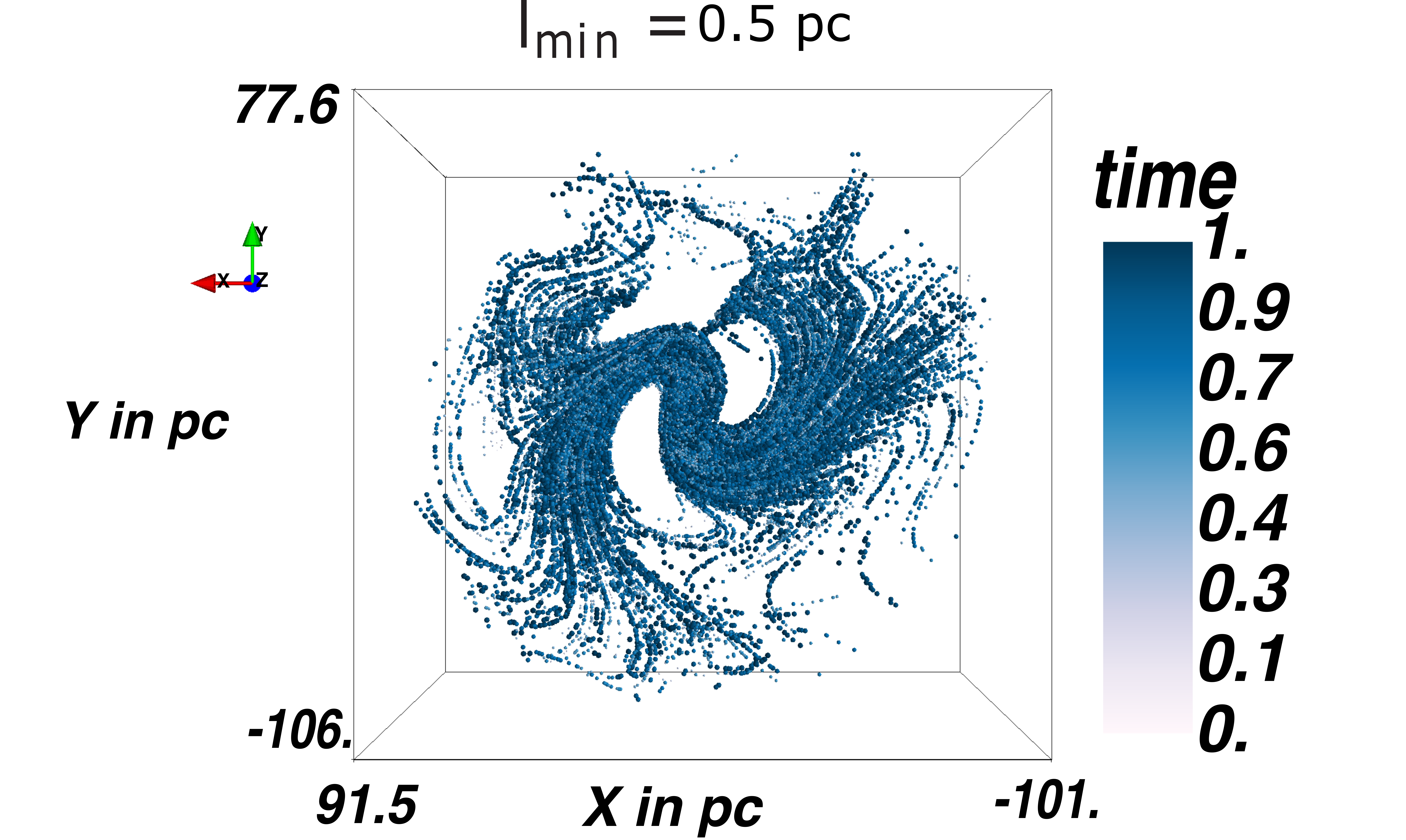}}
\subfigure{\includegraphics[width=0.45\linewidth]{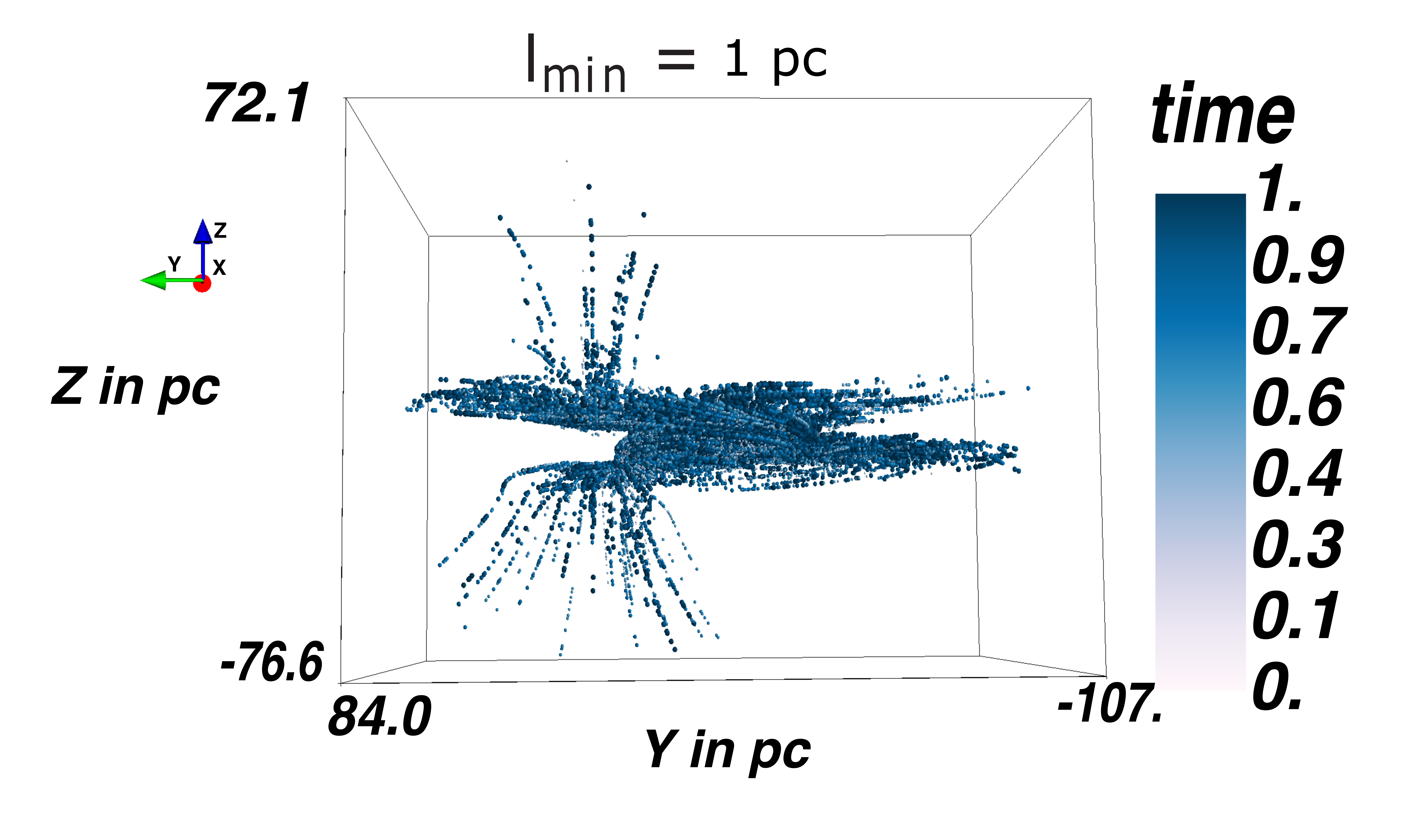}}
\subfigure{\includegraphics[width=0.45\linewidth]{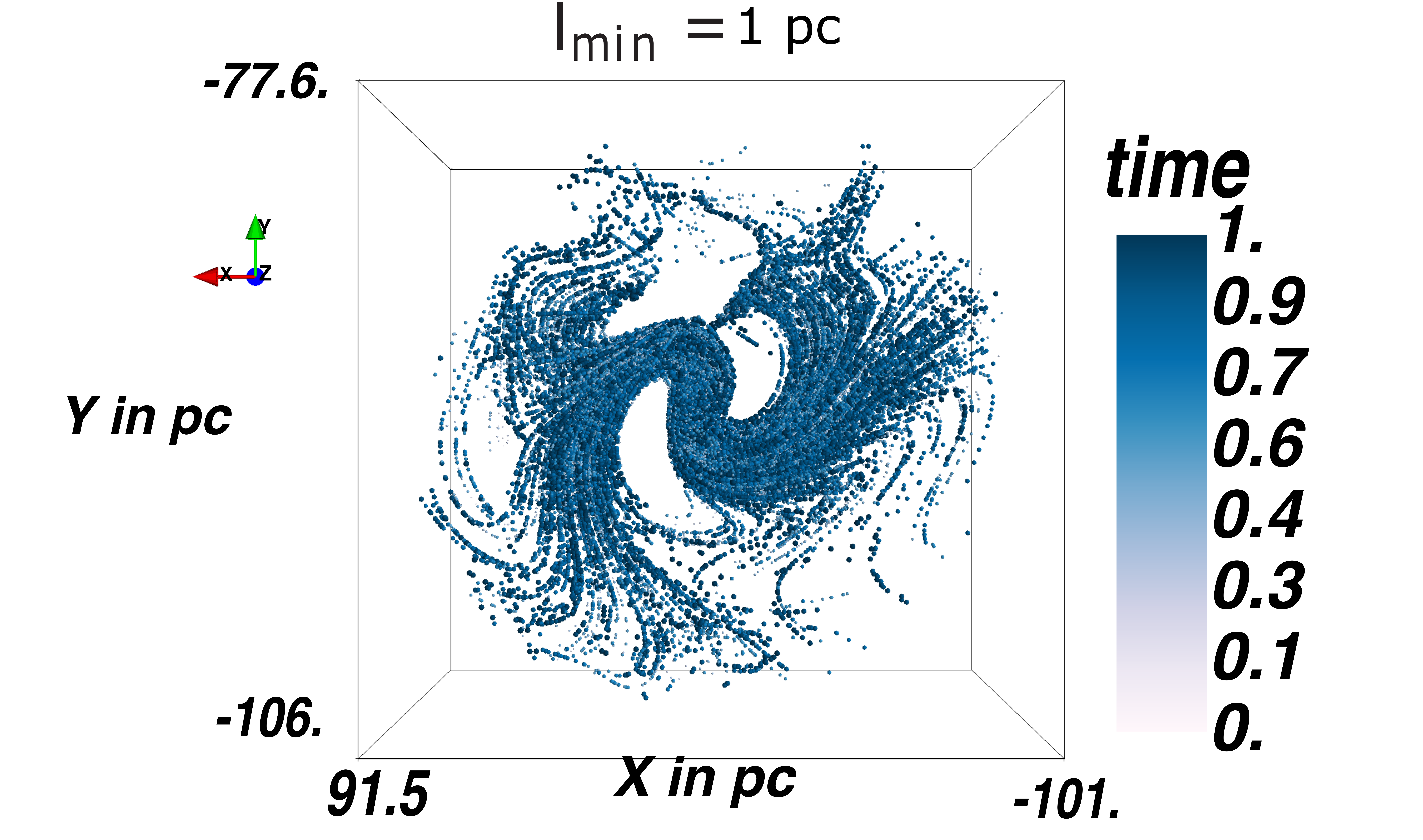}}
\subfigure{\includegraphics[width=0.45\linewidth]{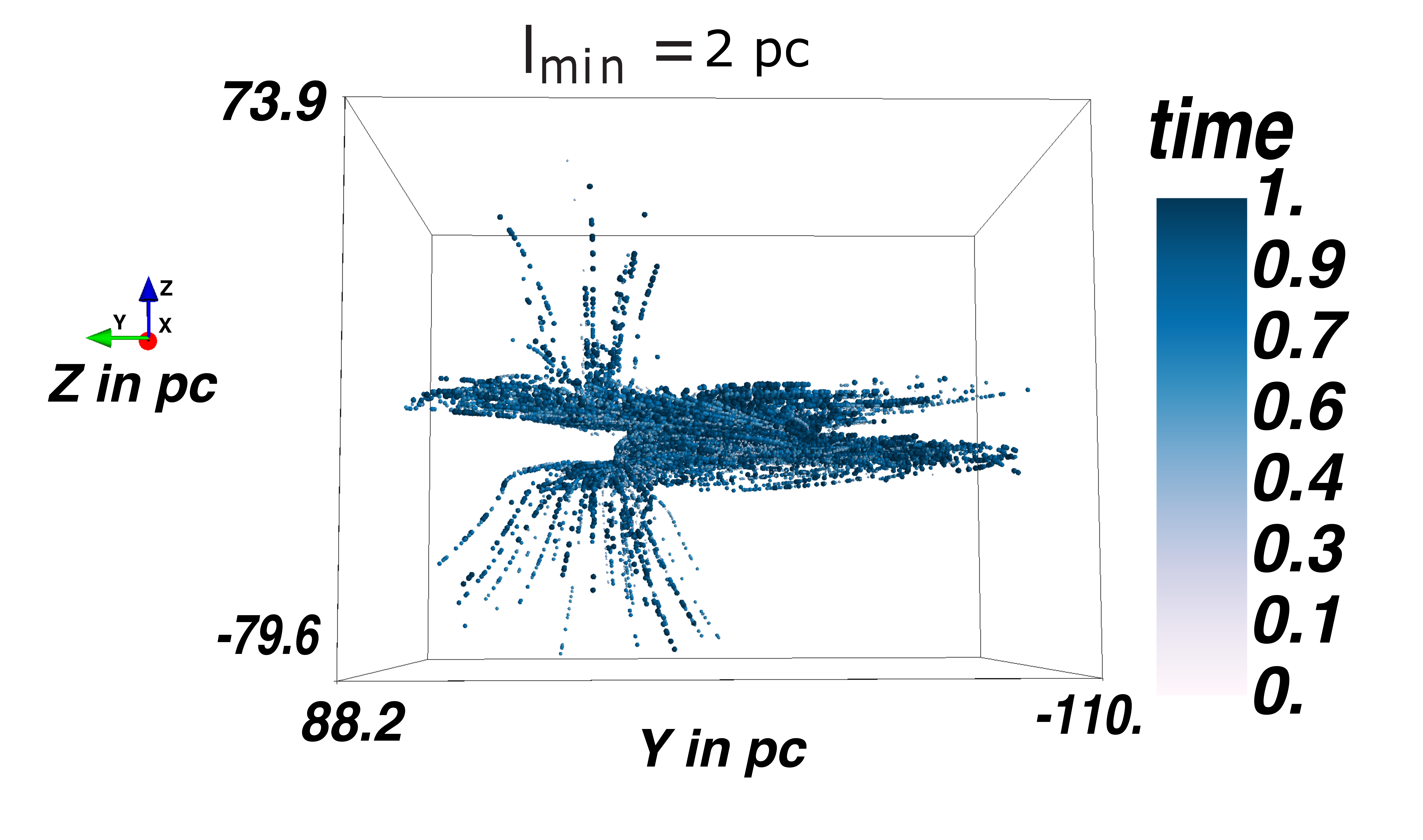}}
\subfigure{\includegraphics[width=0.45\linewidth]{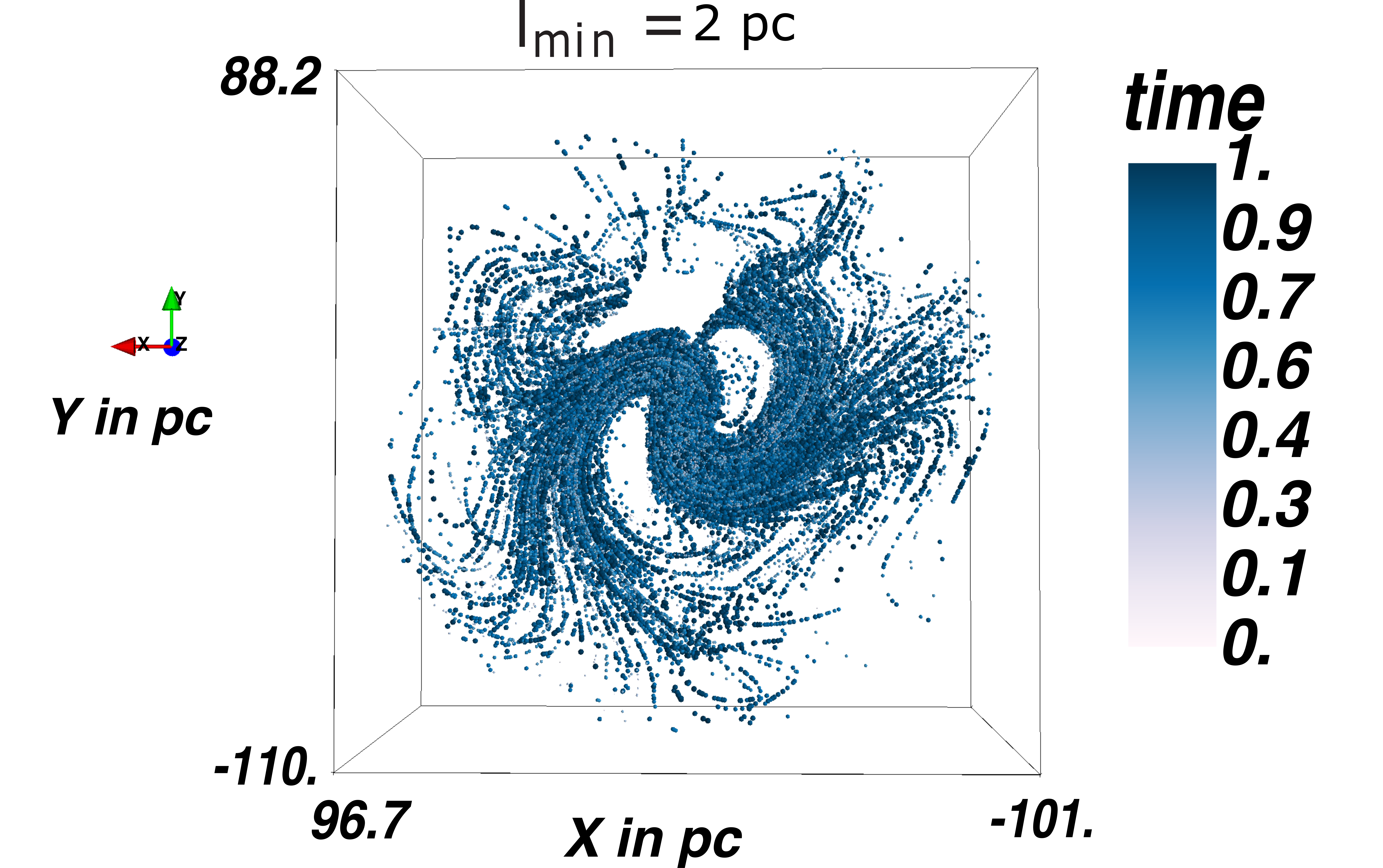}}
\caption{The CR trajectories of $10^4$ protons which were isotropically injected from a centralized source and propagated in the GC magnetic field according to \citep{guenduez_bfield2020}. The time passed since injection is represented by the blue-colored opacity scale.}
\label{fig:Trajectory}
\end{figure*}

%========================================================
In order to optimize the choice of the minimum step length, test simulations are performed with $10^{4}$ protons of $1$~TeV energy, using $\epsilon=0$. Figure \ref{fig:Trajectory} shows test particle trajectories for minimum step lengths of $l_{\min}$/pc=$0.1,\,0.5,\,1.0,\,2.0$, respectively, going from top to bottom. 
We use this interval to be sensitive to the smallest structures in the simulation, which are the smallest MCs, in the order of $0.1-1.7$~pc.  
The left panel shows the $Y-Z$ distribution, i.e.\ the edge-on view of the GC, whereas the right panel shows the $X-Y$ distribution, i.e.\ the face-on view of the GC. The particles are injected directly into the GC and are propagated in the magnetic field configuration presented in \cite{guenduez_bfield2020}, with a detection taking place every $1$~pc. As the smallest MC is on the order of $0.1$~pc, this is the smallest minimum step size that we consider. As the distributions do not show any significant deviations, we use the large value $l_{\rm step, min}=1$~pc in order to save computational time.

In order to determine a useful maximum step length, the poloidal magnetic field in the ICM region is considered as it shows the least curvature and thus allows for the largest steps.
	While in these regions the magnetic field is quite uniform at a length of $\sim 200$~pc, the requirement of at least 25 steps before a particle leaves the simulation
volume provides better reliability. This reduces the step length to $8$~pc and yields $l_{\rm step, max}= 5.0$~pc. With the implementation of the SDE Module in CRPropa3.1 \citep{CRpropa2017} the step size is chosen adaptive to ensure small interaction probability in each step and keep the position error due to the magnetic field integration below $10^{-4}$ per step. 
% In general, in our simulations, smaller values are chosen automatically, if necessary.

%================================================================
\subsection{Source distribution}
\label{SourceDistribution}
%================================================================
Sources in the GC that have the potential to accelerate particles up to knee energies or beyond are:
\begin{enumerate}
	\item
	The central \textbf{supermassive black hole \SgrA\ } located at $l=359.94^{\circ}$ and $b=-0.046^{\circ}$ \citep{SgrACenterMW1998,Petrov2011};

	\item
	The \textbf{SNR Sgr A East} with its center at a distance of 2 pc from \SgrA\ \citep{BlackholeBook};

	\item
	The \textbf{SNR G0.9+0.1} at a distance of $\sim$130 pc from \SgrA\ \citep{HESS2018GC};
	
	\item
	\textbf{other SNRs} such as G359.0 -0.9 \citep{LaRosaFilaments}, G359.10 -0.5 \citep{LaRosaFilaments}, G0.30 +0.04 \citep{G0.30+0.04SNR,LaRosaFilaments}, G0.9 +0.1 \cite{HESS2018GC} and Sgr D \citep{SgrD};

	\item
	The population of \textbf{unidentified pulsars} that could contribute to the gamma-ray excess in the GC observed in the \textit{Fermi}-LAT energy range\footnote{The pulsar (PSR) J1746-285 has already been detected in the H.E.S.S. energy range \citep{HESS2018GC} and is consistent with emission from the pular-wind nebular (PWN) G0.13-0.11.}.
\end{enumerate}
%-----------------------------------------------
\subsubsection{The central source Sgr~A$^{*}$}
%-----------------------------------------------
The center of our Galaxy \SgrA\ is estimated to have a mass of $4.3\cdot 10^6\, M_{\odot}$. The supermassive black hole at the position \SgrA\ has recently been detected directly \citep{gc_smbh2022} and has a Schwarzschild radius of $\sim1.3\cdot10^{12}$~cm \citep{Gilessen2017}. \SgrA\ was first discovered in the radio survey by \cite{Piddington1951} as a discrete and very luminous source. The radio emission of \SgrA\ at $>1$~GHz exhibits a power-law spectrum with a constant spectral index of 1/3, indicating an optically thin synchrotron emission.
It was not until about decades later that higher resolution instruments at $\lambda=3.5$~mm exhibited the elliptical shape of \SgrA\ with a radius of less than $\left( 1/3\cdot 10^{-8}\right) ^{\circ}$ \citep{Rogers1994,Krichbaum1998}, which accordingly allows only an ultra-compact source, i.e. a SMBH. However, the exceptionally low average rate of mass accretion indicates that the SMBH is no longer active today \citep{SgrAinactive}. It is possible that part of the energy of the SMBH is converted into the acceleration of cosmic rays, which makes \SgrA\ to one of the prime candidates concerning the origin of Galactic cosmic rays.
%-----------------------------------------------
\subsubsection{SNRs and other point sources}
%-----------------------------------------------
From the gamma-ray measurements in the GC by H.E.S.S., three point sources have been identified \citep{HESS2018GC}:
\begin{enumerate}
	\item HESS J1745-290 (central source)
	\item G0.9+0.1 (SNR) 
	\item HESS J1746-28 (PSR)
\end{enumerate} 
It is discussed in \citep{HESS2018GC} that the scenario of a steady-state emission fits the profile better than a burst scenario. The steady-state emission is what we investigate in this paper, as compared to the H.E.S.S.\ results, we model the three-dimensional transport in the GC region rather than applying a one-dimensional, simplified model. The scenario of three emitting sources is therefore important to test even for our three-dimensional propagation modeling in order to quantify if even here, the three-source scenario is the best fit or not.
%-----------------------------------------------
\subsubsection{Pulsar population \label{pulsars:sec}}
%-----------------------------------------------
Measurements at GeV energies give hints towards cosmic-ray populations in the GC: An approximately spherically symmetric gamma-ray excess with an extent of 20$^{\circ}$ from the GC has been found by many groups using data from the \textit{Fermi}-LAT (for details, see \cite{Ackermann2017,GC_DiMauro_2021} and references therein). For this purpose, a variety of interstellar emission models and point source catalogs have been studied. Interestingly, the spectral behavior of the excess resembles the expected behavior of PSR emission in the \textit{bulge} of the Galaxy. Thus, one expects that the origin of the \textit{Galactic Center excess} is based on unresolved PSRs. Motivated by this fact, this work makes use of the PSR distribution of the inner Galaxy derived by \cite{MPSRFermi} and obtained by extracting PSR candidates from \textit{Fermi}-LAT data. \cite{MPSRFermi} assume a spherically symmetric description and identify a spatial distribution following a radial profile of $\mathrm{d}n/\mathrm{d}r\propto r^{-\alpha}$ with $\alpha=2.6$ for the \textit{bulge}, i.e. $r<3$~kpc. Further, it has been argued that 800 - 3600 PSRs are required to explain the \textit{Galactic Center excess}. As the distribution of pulsars is important for GeV gamma rays, we will test their influence in our simulations.

\begin{figure}[tb]
	\centering
	\subfigure{\includegraphics[width=0.8\linewidth]{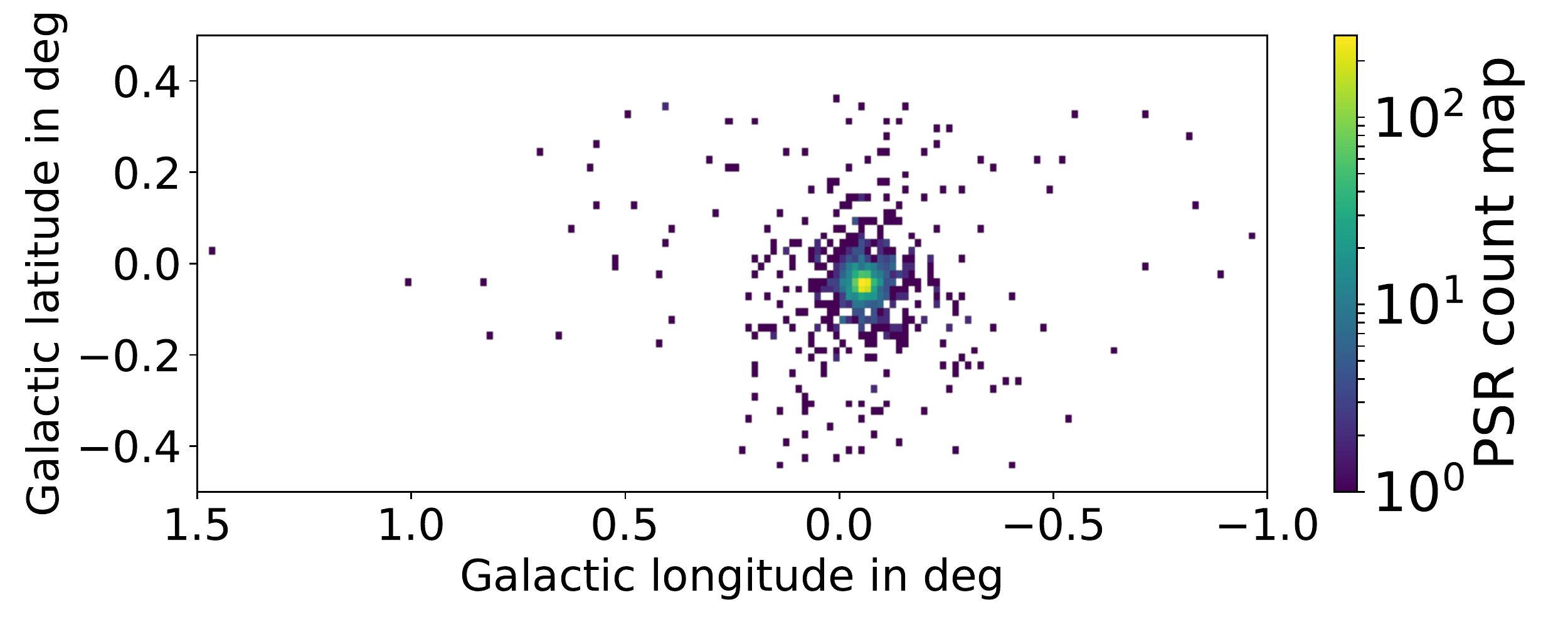}}
	\caption[Distribution of unresolved pulsars]{The expected distribution of unresolved PSRs to explain the \textit{Galactic Center excess} is displayed.}
	\label{fig:PSRPosition}
\end{figure}

In order to include the pulsar distribution in our simulations, we use the following measures:
We normalize the distribution to $1$ for the integrated distribution function within $3$~kpc. In doing so, the distribution of the source density per radius is described as
\begin{equation}
    \frac{\mathrm{d}n}{\mathrm{d}r}=\frac{3-\alpha}{4\pi\, r_{\mathrm{max}}^{3-\alpha}} \cdot r^{-\alpha}, \ \text{for } r<r_{\mathrm{max}}\,.
\end{equation}
Accordingly, the total number of PSRs on the surface $4\pi\,r^2$ per unit length at a specific radius yields
\begin{equation}
    N(r)=4\pi\,r^2 \frac{\mathrm{d}n}{\mathrm{d}r}=\underbrace{\frac{3-\alpha}{{r_{\mathrm{max}}^{3-\alpha}}}}_{\substack{=:\beta}} \cdot r^{2-\alpha},\ \text{for } r<r_{\mathrm{max}}.
\end{equation}
As an input for the simulation, what is needed is the number of pulsars for a given spatial coordinate. Thus, the function $N(r)$ needs to be inverted to find the radius as a function of the source density. The inverse function must ensure that $N^{-1}(N(r))=r$ and $N(N^{-1}(u))=u$ at which $u$ is some particular value of $N(r)$ at a specific radius. In doing so, the inverse function becomes
\begin{equation}
    N^{-1}(u)=\left( \frac{u}{\beta}\right)^{1/(2-\alpha)} \, .
\end{equation}
Assuming that the PSR distribution extends from the \textit{bulge} edge to the center, this range leads to $u\in[0.000129,0.016]$. In order to obtain the position in Cartesian coordinates, one must reduce the number of free parameters due to the spherically symmetric description. Because the CMZ is approximately symmetrical to the $x$-axis, in the next step, the unresolved PSRs are assumed to be located at the densest gas position, i.e. $x=0$. Then, the transformation from spherical to Cartesian coordinates leads to 
\begin{equation}
    y=r\cdot\sin(\theta)= N^{-1}(u) \cdot\sin(\theta) \text{ and } z=N^{-1}(u) \cdot\cos(\theta)\, .
\end{equation}
The positions of PSRs are then obtained by drawing a random number from a uniform distribution of $u\in[0.000129,0.016]$ and $\theta_1\in[-\pi,\pi]$. 
Random numbers are obtained by using the python library \textit{numpy} \citep{Numpy}.
Once a position is obtained, the same procedure can be repeated until the expected average number of 2200 PSRs is reached. 
This setting leads to the PSR distribution within the CMZ region displayed in Figure \ref{fig:PSRPosition}.

%-----------------------------------------------
\subsubsection{Scenarios tested in this work}
%-----------------------------------------------
With the potential sources of cosmic rays in the GC region discussed above, we test five different source setups:
\begin{enumerate}
	\item
	\textbf{[{Sgr~A}$^{\ast}$]} A centralized source at the position of \SgrA, also known as HESS J1745-290, is used as a sole source.

	\item
	\textbf{[{3sr}]} \SgrA, G0.9+0.1 (SNR), and HESS J1746-285 (PSR) are considered as point sources as observed by \cite{HESS2018GC}. 
	Here, we inject $72\%$ of all particles into the central source HESS J1945-290, 22\% in G0.9+0.1 and 6\% in J1746-285. These numbers are based on the H.E.S.S.\ findings of the one-dimensional modeling presented in \cite{HESS2018GC}.

	\item
	\textbf{[{3sr+uPSR}]} In addition to the second source setup, this setting adds the PSR distribution that is discussed in Section \ref{pulsars:sec}. In this scenario, we weight the pulsars to contribute with $19\,\%$ to the total flux, consistent with the weighting factor necessary to explain the \textit{Galactic Center excess} at GeV gamma-ray energies presented in \citep{Ackermann2017}.

	\item
	\textbf{[{uPSR}]} The unresolved PSR distribution is assumed to be the only source contribution.

	\item
	\textbf{[{hom}]} As a test, a cylindrically symmetric and homogeneous source distribution, which includes all known sources, MCs and star clusters is considered. The homogeneous distribution is limited to a cylinder with a radius of $-0.3^{\circ}<b<0.3^{\circ}$ and a length of $-0.5^{\circ}<l<1.1^{\circ}$. While this scenario is not motivated by any underlying physics, it serves as a test and a comparison to the other scenarios that are motivated as described above.
\end{enumerate}
\begin{figure}[tb]
	%	\vspace*{-1cm}
	\centering
	\subfigure{\includegraphics[width=0.9\linewidth]{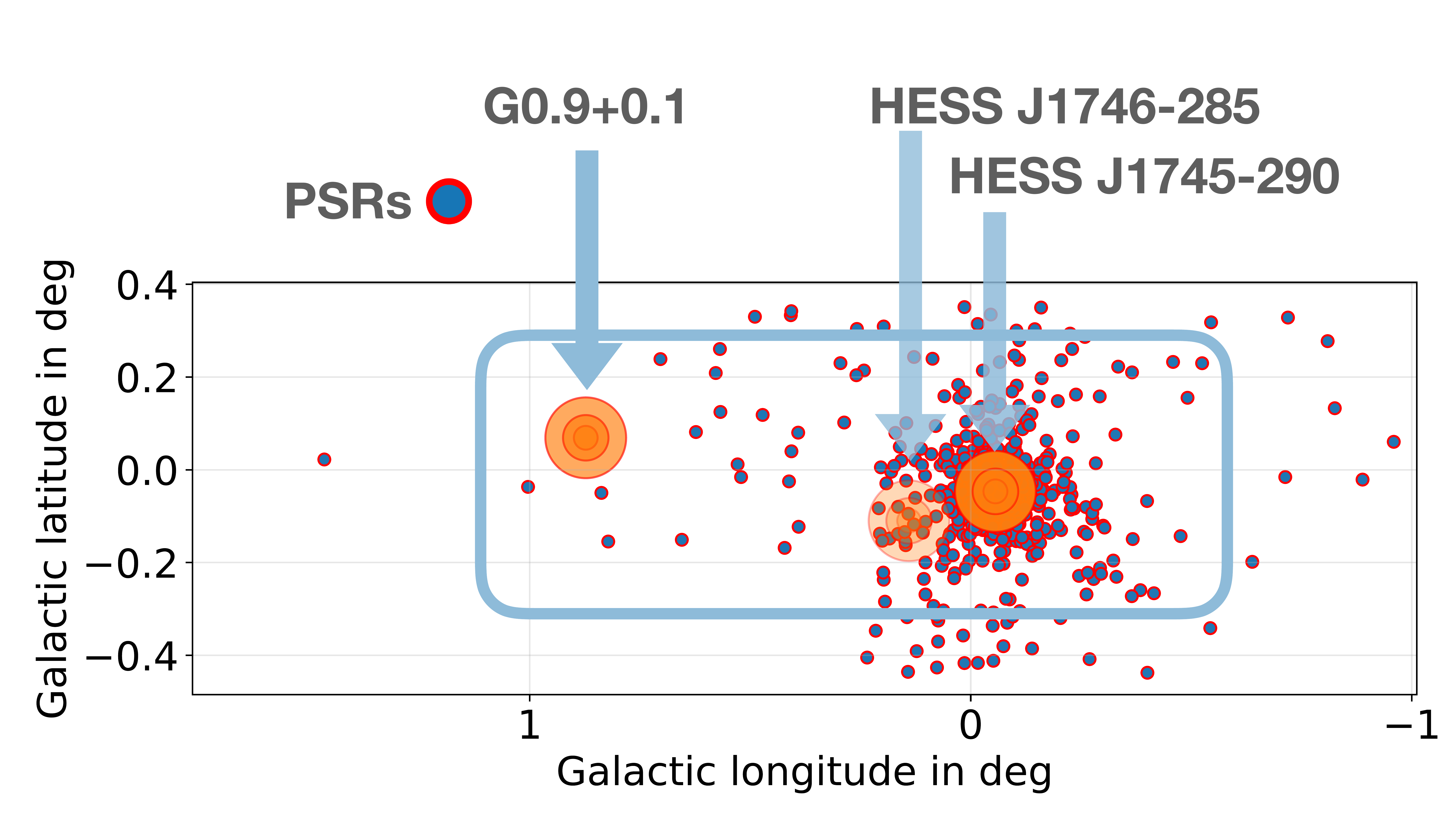}}
	\caption[High-energy source setups for simulations]{High-energy sources contained in all five source setups in the GC are presented as a function of Galactic coordinates. The opacity of the three sources is proportional to the source strength.}
	\label{fig:SourceDistribution}
\end{figure}
Hereafter, the first setup is referred to as  \textbf{[SgrA$^{\ast}$]}, second \textbf{[3sr]}, the third as \textbf{[3sr+uPSR]}, the fourth as \textbf{[uPSR]}, and the fifth as \textbf{[hom]}. All sources are visualized in Fig.\ \ref{fig:SourceDistribution}. Here, the blue frame represents the cylindrically symmetrical source distribution, the red-edged and blue-filled circles represent the PSRs, and the red-edged and orange filled circles represent HESS J1745-290, G0.9+0.1 and HESS J1746-285, respectively. Furthermore, the opacity of the orange-filled sources is proportional to the source strength.

%====================================================
\subsection{Tested particle population and energy range}
%====================================================
In this simulation, for simplicity, we use protons as the primary cosmic-ray component. We neglect the contribution from the elements of helium and higher mass numbers at this point, as the error is expected to be rather small, as in the energy range of interest, the flux is dominated by protons, see e.g.\ \cite{beckertjus2020} for a review.

The minimum energy is set to the lower threshold energy that applies to \textit{CRPropa}, which is written for ultra-relativistic particles in the limit $E\sim p\cdot c$ only. For protons, this results in a minimum energy of $E_{\min}=1$~TeV. This means that the secondary photon flux that can be considered must have a lower limit in energy $>100$~GeV. We therefore restrict the interpretation of our results to the H.E.S.S.\ data and refrain from comparing to \textit{Fermi}-LAT data at this point. 

The maximum energy is assumed to correspond to the maximum expected Galactic CR energy of $1$~PeV, i.e. the energy at the \textit{knee}. The spectral index is taken from the results of the theory of stochastic acceleration in shock vicinities and therefore set to $\alpha=2.0$ for the simulations. As the SDE method allows for the reweighting of the results, this spectral index can be varied in the analysis in order to find the best fit.

%================================================================
\subsection{Interactions}
\begin{figure}[ht]
	%	\vspace*{-1cm}
	\centering
	\subfigure{\includegraphics[width=1.0\linewidth]{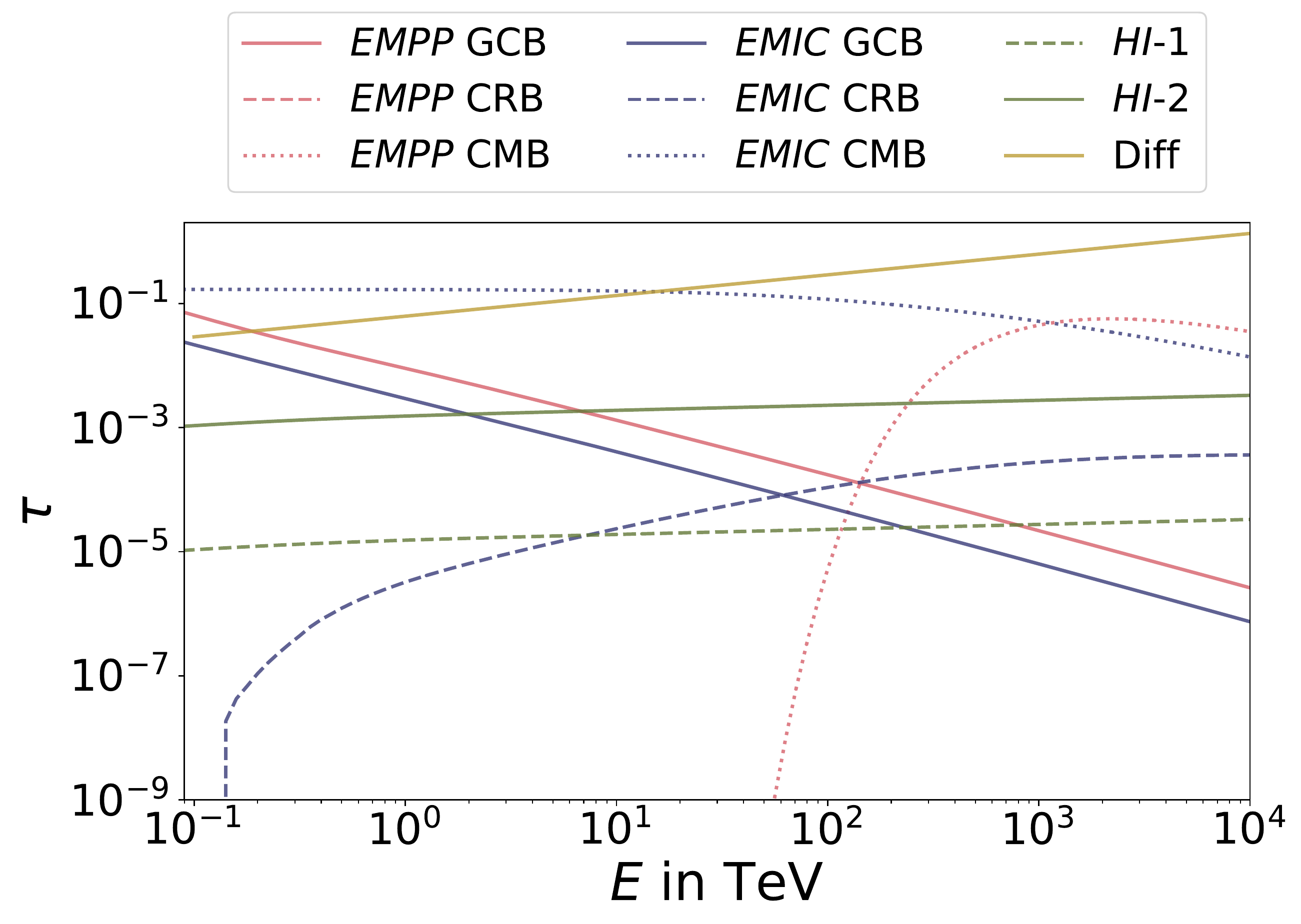}}
	\caption{The optical depth is presented as a function of the energy and for different interaction processes: diffusion (yellow, thick solid line), Inverse Compton in blue for the CRB (dotted), GCB (short dot-dashed) and CMB (long dot-dashed), pair-production in red for the CRB (short dashed), GCB (thin solid) and CMB (long dashed) as well as hadronic interactions in green with target densities of $10^{2}$~cm$^{-3}$ and $10^{3}$~cm$^{-3}$ (short and long dot-dot-dashed), respectively. Here, a region with a radius of 200~pc is considered and $\lambda$ denotes the mean free path, which is correlated with the loss timescale.}
	\label{fig:OpticalDepth}
\end{figure}

%================================================================
The GC is enhanced in gas density and bright in background photons over a wide range of wavelengths. This abundance of background gas and photons induces different processes, i.e.\ hadronic interactions, electromagnetic pair production and inverse Compton scattering. These processes are taken into account in the numerical simulation approach. The related modules in \textit{CRPropa} are called \textit{HI} (hadronic interactions), \textit{EMPairProduction}, \textit{EMPP} (electromagnetic pair production), and \textit{EMInverseComptonScattering}, \textit{EMIC}  (electromagnetic inverse Compton scattering). The module \textit{HI} requires an input function concerning the gas density distribution. \textit{EMPairProduction} and \textit{EMInverseComptonScattering} have equivalent input parameters that require the photon density as a function of the wavelength and a scaling-grid obtained from the spatial distribution of the photon field. In addition to the GC background photon field presented in Section \ref{PhotonField} (from now on called GCB), CMB and Cosmic Radio Background (CRB) \citep{1996Protheroe} radiation are considered. The latter two are homogeneously distributed and already included in the public version of \textit{CRPropa 3.1.5}. All interaction modules are labeled so that secondary particles can be traced back to the generation process or the photon field. The corresponding average optical depths indicating the relevance of a particular process are presented in Figure \ref{fig:OpticalDepth}. The optical depth for $HI$ is calculated for a target medium with a gas density of $10^2$~cm$^{-3}$ ($HI-1$) and $10^4$~cm$^{-3}$ ($HI-2$). 

%================================================================
\subsection{Observer and Output}
%================================================================
The \textit{CRPropa} module \textit{Observer} allows to customize the output file. Here, the user is able to construct an observer surface that detects particles only when they cross the surface. The surface may have a paraxial or a spherical shape. However, because this work is focused on secondaries, the output is not restricted to an observer sphere. All secondary particles are recorded immediately after their generation and remain an active part of the simulation.\par
This work takes the following information of the particles, subsequently referred to as \textit{Candidates}, into account: current position, current energy, current direction of motion, trajectory length, serial number, particle type, and particle type of the parent particle, position of their source, and the generation process. Because the generation process is tagged, the gamma-ray attenuation can be considered by subtracting all gamma rays, which scatter into electron-positron pairs via the module \textit{EMPP}.

%==============================================
%\subsection{Normalization of the results}
%==============================================

%%%%%%%%%%%%%%%%%%%%%%%%%%%%%%%%%%%%%%%%%%%%%%%%%%%%%%%%%
\section{Results\label{results:sec}}
%%%%%%%%%%%%%%%%%%%%%%%%%%%%%%%%%%%%%%%%%%%%%%%%%%%%%%%%%
In this section, we present our final simulations with the settings as discussed in Section \ref{propagation:sec}. We compare the results with the H.E.S.S.\ data presented in \cite{HESS2018GC}. 
The gamma-ray measurements by  H.E.S.S.\ pass through many internal conditions and filters, such as the effective area, which is in general sensitive to energy. In this particular case, however, the considered observed data are not expected to change with respect to the energy dependence of the effective area, since the position of the GC in the sky relative to the H.E.S.S. telescopes is approximately at the zenith. Therefore, the observation of the GC can be achieved under the best conditions with a low energy threshold and without any significant change in the effective area through the entire CMZ \citep{Benbow2005}. We therefore do not have to correct for any effects concerning the detection efficiency of H.E.S.S.

%================================================================
\subsection{Spatial gamma-ray count profiles}
\begin{figure*}[hp]
% 	%	\vspace*{-1cm}
% 	\centering
% 	% SgrA*
% 	\subfigure{\includegraphics[width=0.42\linewidth]{figures_using/FindEpsilon/NProfileLat_1Sr_1e+06_Emax=1e+15eV.pdf}}
% 	\subfigure{\includegraphics[width=0.42\linewidth]{figures_using/FindEpsilon/NProfileLong_1Sr_1e+06_Emax=1e+15eV.pdf}}
% 	% 3sr
% 	\subfigure{\includegraphics[width=0.42\linewidth]{figures_using/FindEpsilon/NProfileLat_3Sr_1e+06_Emax=1e+15eV.pdf}}
% 	\subfigure{\includegraphics[width=0.42\linewidth]{figures_using/FindEpsilon/NProfileLong_3Sr_1e+06_Emax=1e+15eV.pdf}}
%     % 3sr+uPSR
% 	\subfigure{\includegraphics[width=0.42\linewidth]{figures_using/FindEpsilon/NProfileLat_3+mpsrSr_1e+06_Emax=1e+15eV.pdf}}
% 	\subfigure{\includegraphics[width=0.42\linewidth]{figures_using/FindEpsilon/NProfileLong_3+mpsrSr_1e+06_Emax=1e+15eV.pdf}}
% 	% uPSR
% 	\subfigure{\includegraphics[width=0.42\linewidth]{figures_using/FindEpsilon/NProfileLat_mpsrSr_1e+06_Emax=1e+15eV.pdf}}
% 	\subfigure{\includegraphics[width=0.42\linewidth]{figures_using/FindEpsilon/NProfileLong_mpsrSr_1e+06_Emax=1e+15eV.pdf}}
% 	% hom
% 	\subfigure{\includegraphics[width=0.42\linewidth]{figures_using/FindEpsilon/NProfileLat_cylSr_1e+06_Emax=1e+15eV.pdf}}
% 	\subfigure{\includegraphics[width=0.42\linewidth]{figures_using/FindEpsilon/NProfileLong_cylSr_1e+06_Emax=1e+15eV.pdf}}
\centering
\includegraphics[width=.92\textwidth]{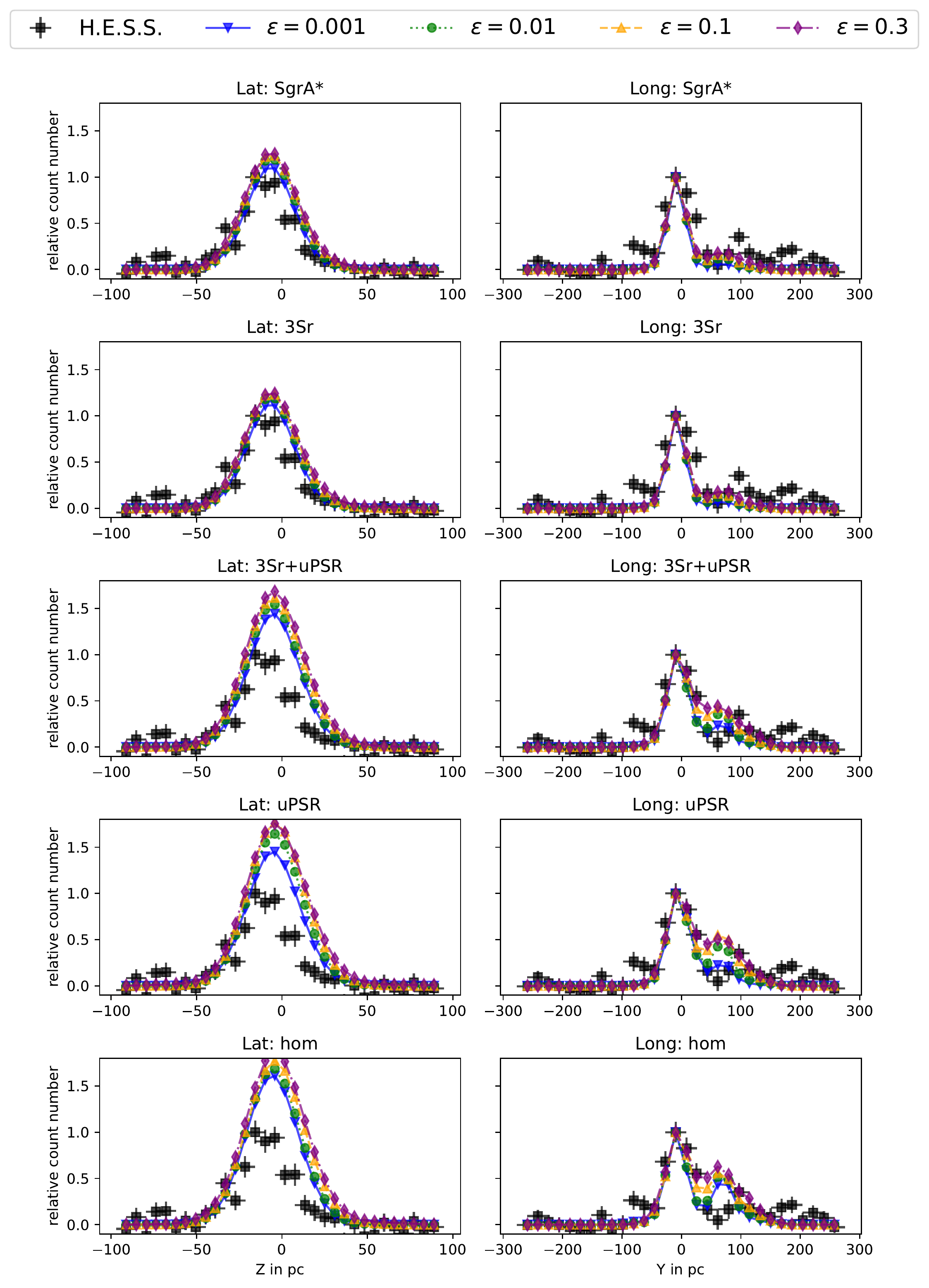}
	\caption{Relative latitudinal (left) and longitudinal (right) count profile in which the counts have been integrated over $|l|<0.5^{\circ}$ and $|b|<0.3^{\circ}$, respectively. Each row shows a simulation with a different source distribution. For reference, the H.E.S.S. observation is displayed, and the background large scale component, as described in \cite{Abramowski2014}, is subtracted.}
	\label{fig:CountMap}
\end{figure*}

%================================================================
In \cite{HESS2018GC}, the gamma-ray count profiles are given as the number of photons detected in a certain longitudinal or latitudinal interval. In this section, we produce count maps from our simulation data:
The longitudinal profiles are integrated over the latitudes from $-0.3^{\circ}$ to $-0.3^{\circ}$ and the latitudinal profiles over the longitudes from $-0.5^{\circ}$ to $0.5^{\circ}$. Because the ratio of perpendicular to parallel diffusion $\epsilon$ has a significant impact on the latitudinal and longitudinal profile, this parameter can be fixed by comparing simulations of different $\epsilon$ values with the count maps as measured by \cite{HESS2018GC}. In order to reproduce the measurements, the simulation results must further be smeared to the H.E.S.S. angular resolution of 0.077$^{\circ}$\footnote{The angular resolution is achieved by considering a point spread function corresponding to a $68\,\%$ containment radius.}. For this purpose, the results from the simulations are treated with \textit{2D-Gaussian smoothing} with a sigma corresponding to a radius of 0.077$^{\circ}$. Furthermore, the simulated data are normalized to the amplitude value at the center of the longitudinal profile. The significance of the detection is the largest at the center, so that the background noise can be neglected. We use the same binning as H.E.S.S.\ for our simulation data. The simulated latitudinal and longitudinal count maps for five different source distributions and four different $\epsilon$ values in each plot are presented in Figure \ref{fig:CountMap}.
%\clearpage

The uppermost panels show the latitudinal (left) and longitudinal (right) profiles that are expected for the \textbf{[\SgrA]} model, the second row shows the model \textbf{[3sr]}, the third one \textbf{[3sr+uPSR]}, the fourth \textbf{[uPSR]} and the lowest panel \textbf{[hom]}. In each plot, the simulation results are shown for four different $\epsilon$ values, i.e.\ $\epsilon = 0.001,\,0.01,\,0.1,\,0.3$. In each plot, the H.E.S.S.\ data are shown as filled squares.

For all source distributions, $\epsilon=0.01$ and $\epsilon=0.001$ fit slightly better to the observations as compared to the results obtained from $\epsilon=0.1$ and $\epsilon=0.3$ and thus, a low influence of perpendicular diffusion is expected. More importantly, it can be seen that the source distribution has a significant impact on the latitudinal and longitudinal profiles. It appears that only the source distributions \textbf{[\SgrA]} and \textbf{[3sr]} deliver a good description of the data for both the latitudinal and longitudinal profiles. Our results show that the \textbf{[3sr]} scenario provides the best fit. Adding the pulsar distribution does not change the result for the longitudinal profile significantly\footnote{the longitudinal part actually fits the data slightly better for the pulsar distribution}, but the latitudinal profile is enhanced such that the prediction overshoots the data significantly in particular for positive $Z$ values. The reduced source distribution with only the central source, \textbf{[\SgrA]}, narrows down the longitudinal profile instead. For the homogeneous distribution of sources, the latitudinal profile is largely overestimated and the longitudinal profile does not fit well, in particular for positive values of $Y$. It is interesting to note that the three-dimensional propagation model comes to the same conclusion as the one-dimensional approximation used in \cite{AbramowskiNature}, i.e.\ that a three source scenario is the best fit for the data.

Thus, for the energy spectrum and the count maps that we present, we restrict ourselves to show the results for the \textbf{[3sr]} scenario. 

For the interpretation of the longitudinal emission profile, it can be summarized that the complex and asymmetric structure, with a number of smaller peaks around the central one, is difficult and none of the models fit all peaks simultaneously. The first peak is related to the high gas density at the center and all models can reproduce it relatively well. The second observed peak location, however, is shifted with respect to the peak predicted in this work. While the peak in our simulations is caused by the dense gas of the dust ridge clouds, the observed peak rather coincides with the location of Sgr~B2. These findings indicate that the gas density taken from \cite{SgrB2} and used in our  simulations is not compact enough. If the mass that is connected to Sgr~B2 is distributed onto a smaller volume with a higher density, the second peak at the location of Sgr~B2 could be enhanced. For clarification, further observations by instruments such as CTA will provide significantly better spatial resolution and cover a broader energy range. Improvements on the gas density measurements will also help to better quantify the fluxes.\par
The peak at $Y \sim 60$ pc in all source setups is due to the compact MCs dust ridges A - F. Their compactness seems to be sufficient and reasonable for the source setups that provide the best fit, i.e.\ \textbf{[\SgrA]} and \textbf{[3s]}. The enhanced emission at $Y\sim -100$~pc to $Y\sim -30$~pc is not reproduced by this model.

For the future, for a reliable identification of a full source setup, the detailed knowledge on the MCs is crucial. The observed third peak is not seen in the simulation results because no relevant sources or MCs are located between $l\sim-40$ pc and $l\sim-80$. The MC Sgr C at $\sim82$ pc has a radius of approximately 1.7 pc and alone would not be sufficient to explain the third peak. Thus, in the given vicinity, additional MCs or unresolved active sources could be present. Alternatively, the size of Sgr C and thus the mass could be underestimated in the observational results. For a convenient comparison, Tables \ref{tab:chisquareLong} and \ref{tab:chisquareLat} list the resulting chi-square tests adapting the null hypothesis that the observed data are described by the simulation.\par 

To summarize, a more realistic gas distribution is necessary in order to explain all of the features seen in the gamma-ray data. The H.E.S.S.\ collaboration has developed an empirical model for the gas distribution, see \cite{HESS2018GC} and references therein, based on the gamma-ray distribution. This, however implies the knowledge of the source distribution. As both cannot easily be disentangled from each other, increasing the knowledge on the gas distribution itself would be of high importance.

\begin{table*}[ht]
	%\hspace*{-2.2cm}
	\caption[Longitudinal chi-square test]{The longitudinal chi-square test of different source setups is listed.}
	\centering
	\begin{tabular}{||c||c|c|c|c|c||} 
		\hline
		$\epsilon$/Source& $\left[\mathrm{Sgr A}^{\ast}\right]$&$\left[\mathrm{3sr}\right]$&$\left[\mathrm{3sr+uPSR}\right]$&$\left[\mathrm{uPSR}\right]$&$\left[\mathrm{hom}\right]$\\
		\hline\hline
		0.3&2.12&2.16&4.91&6.37&9.37\\
		\hline
		0.1&2.25&2.24&4.62&7.17&7.31\\ 
		\hline
		0.01&2.44&2.48&3.71&4.70&6.50\\
		\hline
		0.001&2.6&2.65&2.56&2.45&5.11\\		
		\hline\hline
	\end{tabular}				
	\label{tab:chisquareLong}
\end{table*}
\begin{table*}[ht]
	%\hspace*{-2.2cm}
	\caption[Latitudinal chi-square test]{The latitudinal chi-square test of different source setups is listed.}
	\centering
	\begin{tabular}{||c||c|c|c|c|c||} 
		\hline
		$\epsilon$/Source& $\left[\mathrm{Sgr A}^{\ast}\right]$&$\left[\mathrm{3sr}\right]$&$\left[\mathrm{3sr+uPSR}\right]$&$\left[\mathrm{uPSR}\right]$&$\left[\mathrm{hom}\right]$\\
		\hline\hline
		0.3&1.83&1.95&11.34&15.41&17.09\\
		\hline
		0.1&1.49&1.444&8.56&12.57&12.64\\ 
		\hline
		0.01&0.93&0.91&5.14&7.79&7.35\\
		\hline
		0.001&0.46&0.52&3.65&4.02&5.46\\		
		\hline\hline
	\end{tabular}				
	\label{tab:chisquareLat}
\end{table*}
\begin{table*}[t]
	%\hspace*{-2.2cm}
	\centering
	\caption[Chi-square test for different spectral indices]{The spectral chi-square test of different spectral indices and $\epsilon$ values is listed.}
	\begin{tabular}{||c||c|c|c|c|c|c|c|c|c||} 
		\hline 
		$\epsilon$/$\alpha$& 2.0&2.05&2.1&2.15&2.2&2.25&2.3&2.35&2.4\\
		\hline\hline
		$0.3$&3.5&3.1&2.8&2.8&2.9&3.3&3.5&4.2&4.7 \\ \hline
		$0.1$&6.4&4.7&3.7&3.0&2.5&2.6&2.6&2.9&3.3\\ \hline
		$0.01$&6.8&4.7&3.6&2.6&2.2&2.0&2.5&2.5&2.5\\ \hline
		$0.001$&8.5&5.9&4.4&3.3&2.5&2.1&1.9&2.1&2.3\\ 
		%$\chi_{\mathrm{A}}^2$&-&-&2.1&-&2.5&-&6.6&-&13.1\\
		\hline\hline
	\end{tabular}				
	\label{tab:chisquareSpectral}
\end{table*}  
%================================================================
\subsection{The energy spectrum}
%================================================================
\label{GammaFlux}
The energy spectrum of the CMZ as measured by H.E.S.S.\ is of particular interest, as it is the first to extend to almost $100$~TeV photon energy. For a hadronic interpretation, that implies the existence of cosmic rays with $\sim$ PeV energies or higher, thus providing the first measurement of a possible \textit{PeVatron} in the Galaxy.

\begin{figure}[t]
	%	\vspace*{-1cm}
	\centering
	\subfigure{\includegraphics[width=0.8\linewidth]{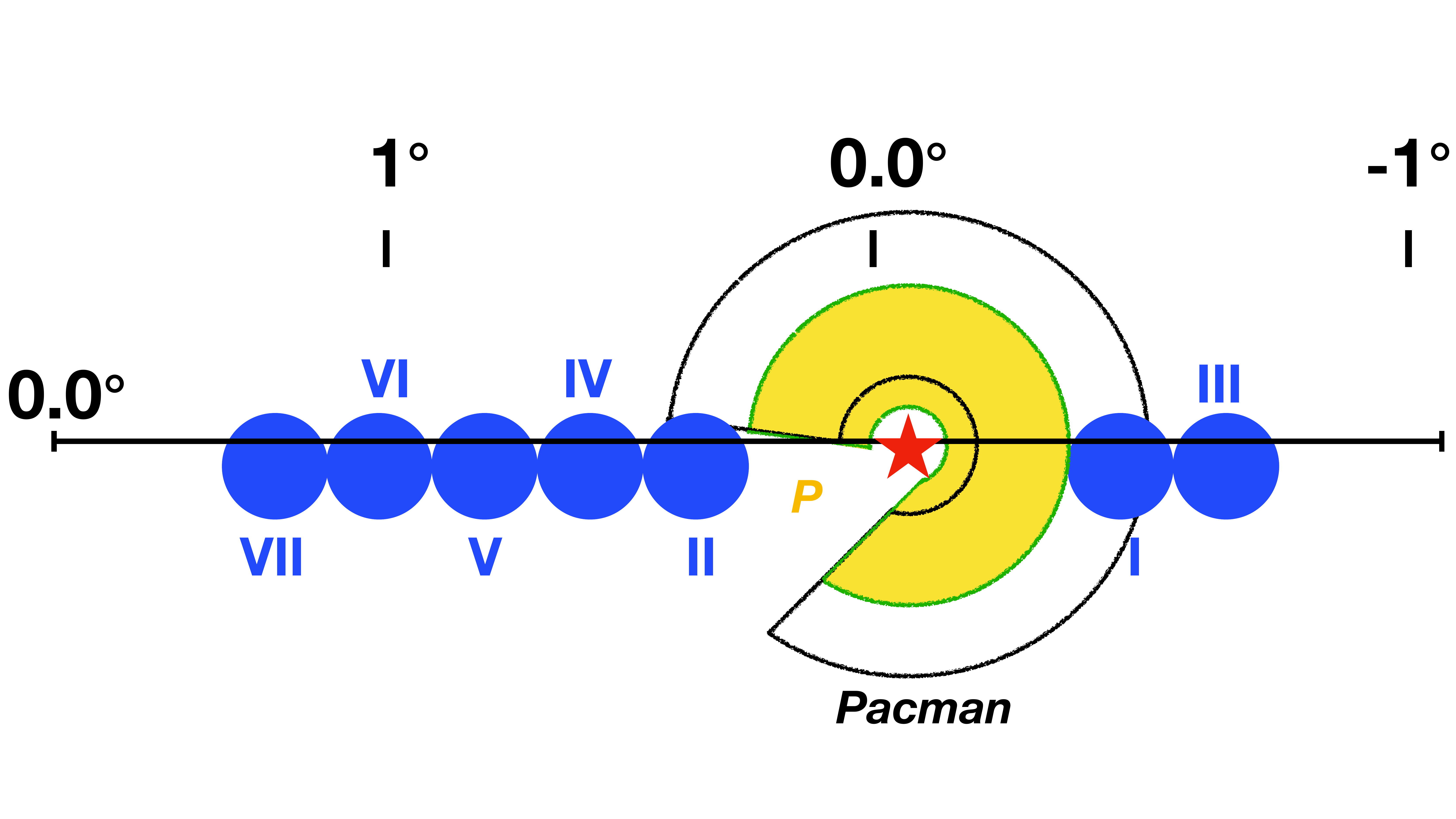}}
	\caption{A schematic view of the regions in the Galactic Center that are used by H.E.S.S.\ to present the spatially resolved signatures. The blue-filled circles have a radius of $0.1^{\circ}$ and correspond to the detection of H.E.S.S. 
	The so-called \textit{Pacman region} is labeled with a yellow $P$. It extends from $0.2^{\circ}$ to $0.3^{\circ}$, where a section $66^{\circ}$ is excluded. The y-axis describes the Galactic latitude $b$ and the x-axis the Galactic longitude $l$ which both are measured in degrees. Considering a distance of $d=8.5$~kpc between the GC and Earth, the \textit{Pacman} region extends approximately up to 70~pc.}
	\label{fig:Regions}
\end{figure}

%%%%%%%%%%%%%%%%%%%%!!!!!!!!!!!!!
The simulation data that we present in the following are integrated over the so-called \textit{Pacman} region, which corresponds to a ring-like, symmetric region of a radius of 0.45$^{\circ}$ around \SgrA, where the inner 0.15$^{\circ}$ and a section of 66$^{\circ}$ are excluded \citep{AbramowskiNature}. A schematic view of the Pacman region together with seven more regions of detection that form the basis of the spatially resolved signal presented in \cite{AbramowskiNature} and \cite{HESS2018GC} are shown in Fig.\ \ref{fig:Regions} .

\begin{figure*}[ht]
	%	\vspace*{-1cm}
	\centering
	\subfigure{\includegraphics[width=0.45\linewidth]{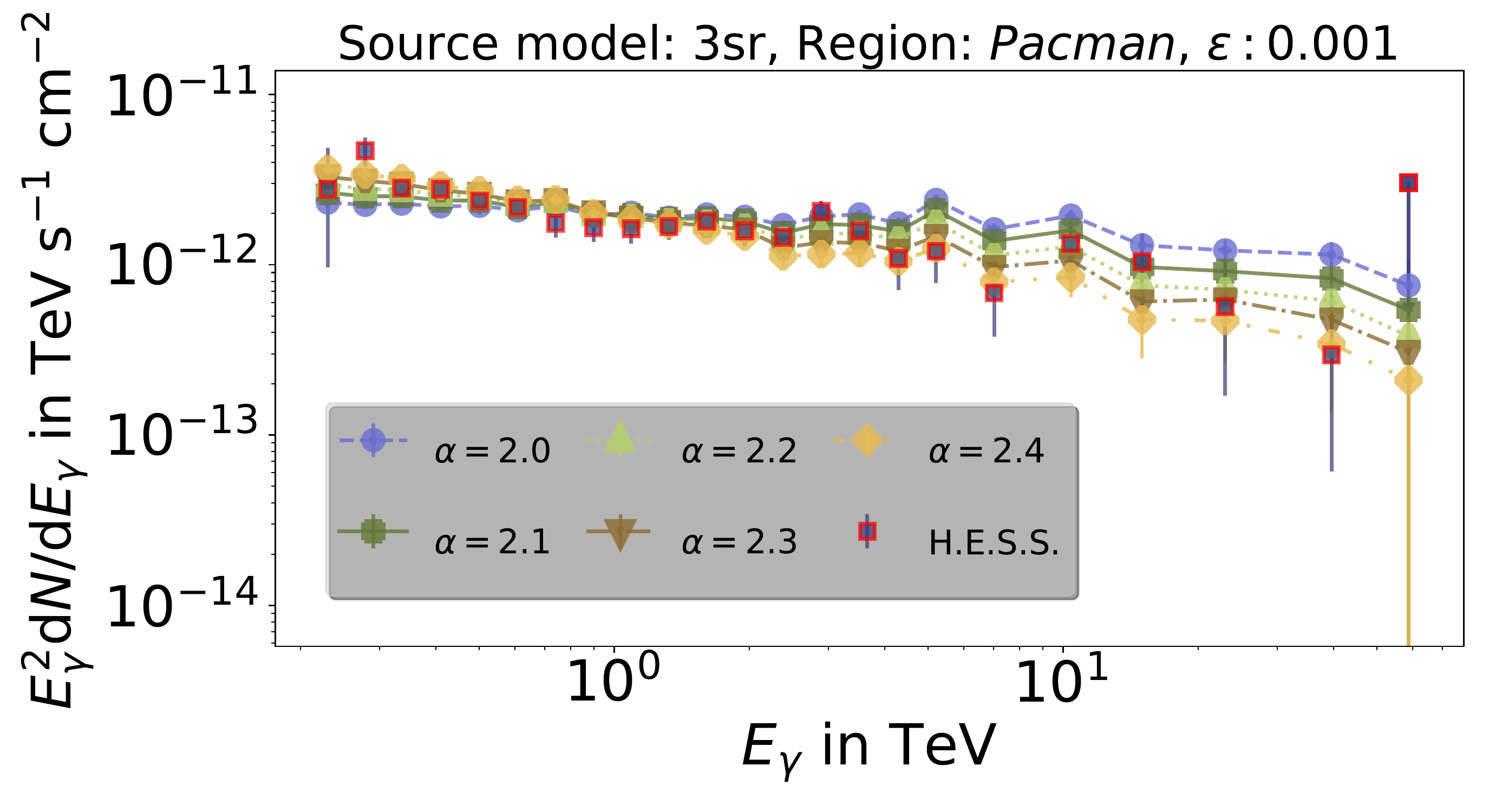}}
	\subfigure{\includegraphics[width=0.45\linewidth]{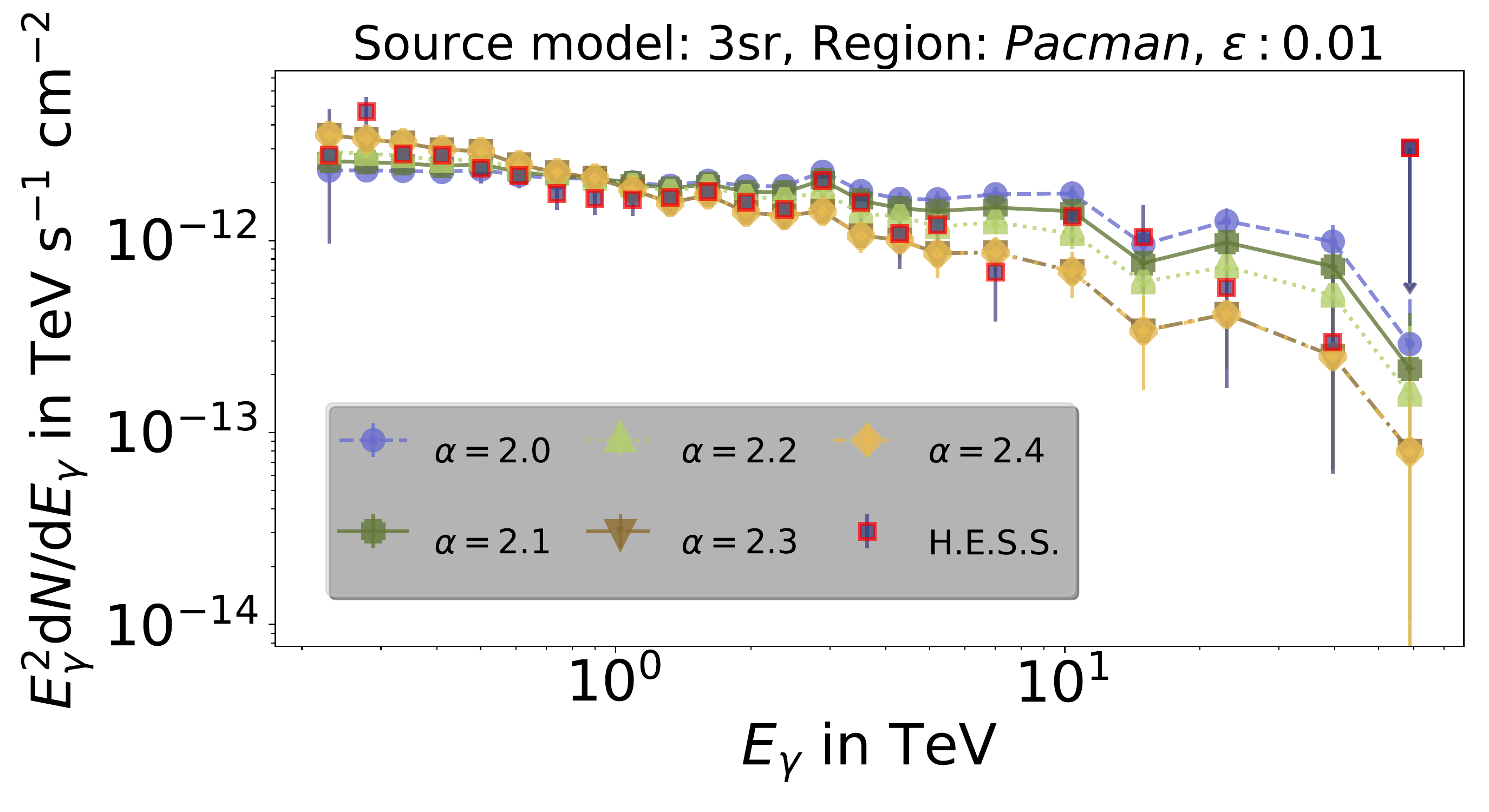}}
	\subfigure{\includegraphics[width=0.45\linewidth]{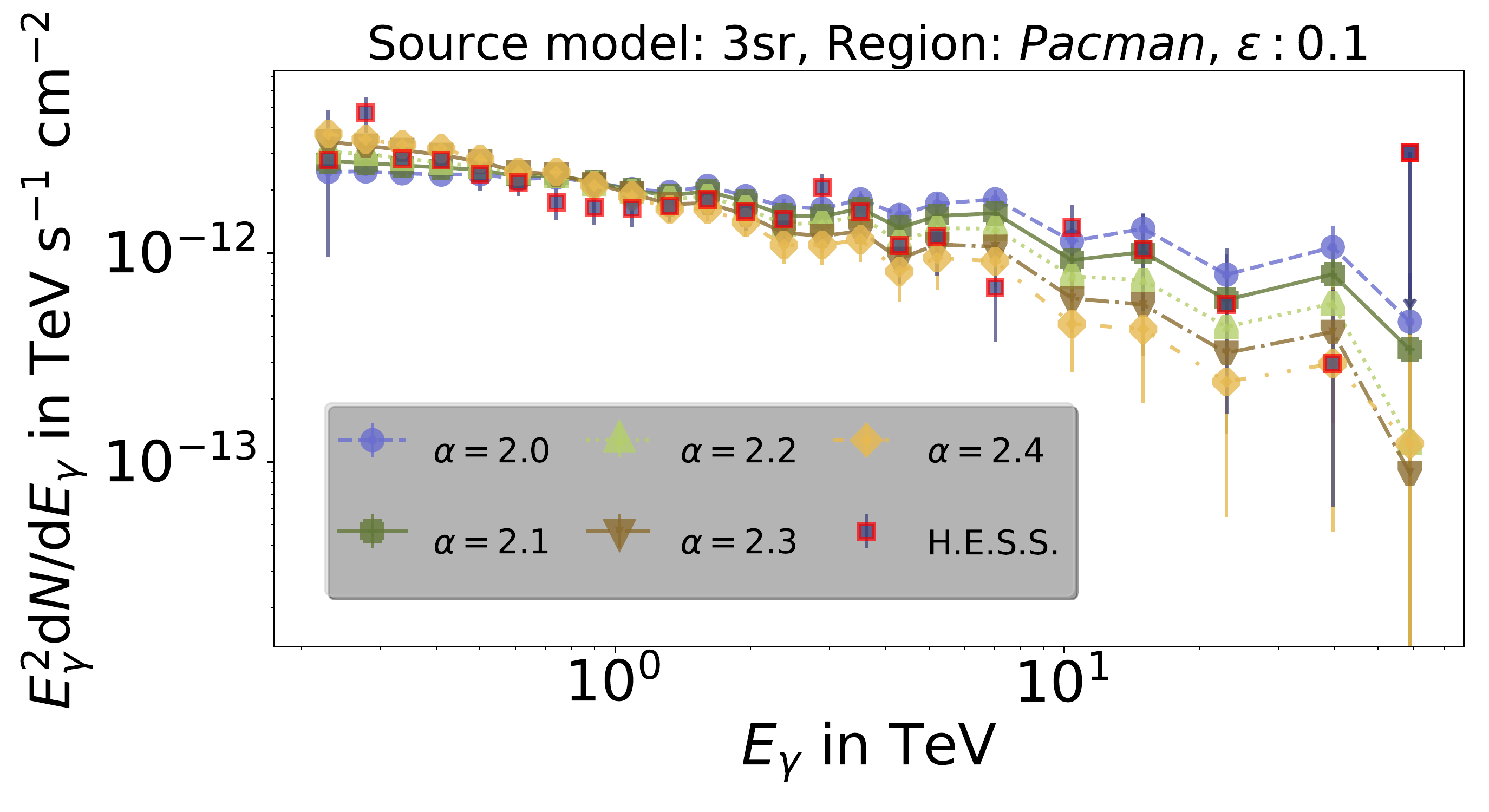}}
	\subfigure{\includegraphics[width=0.45\linewidth]{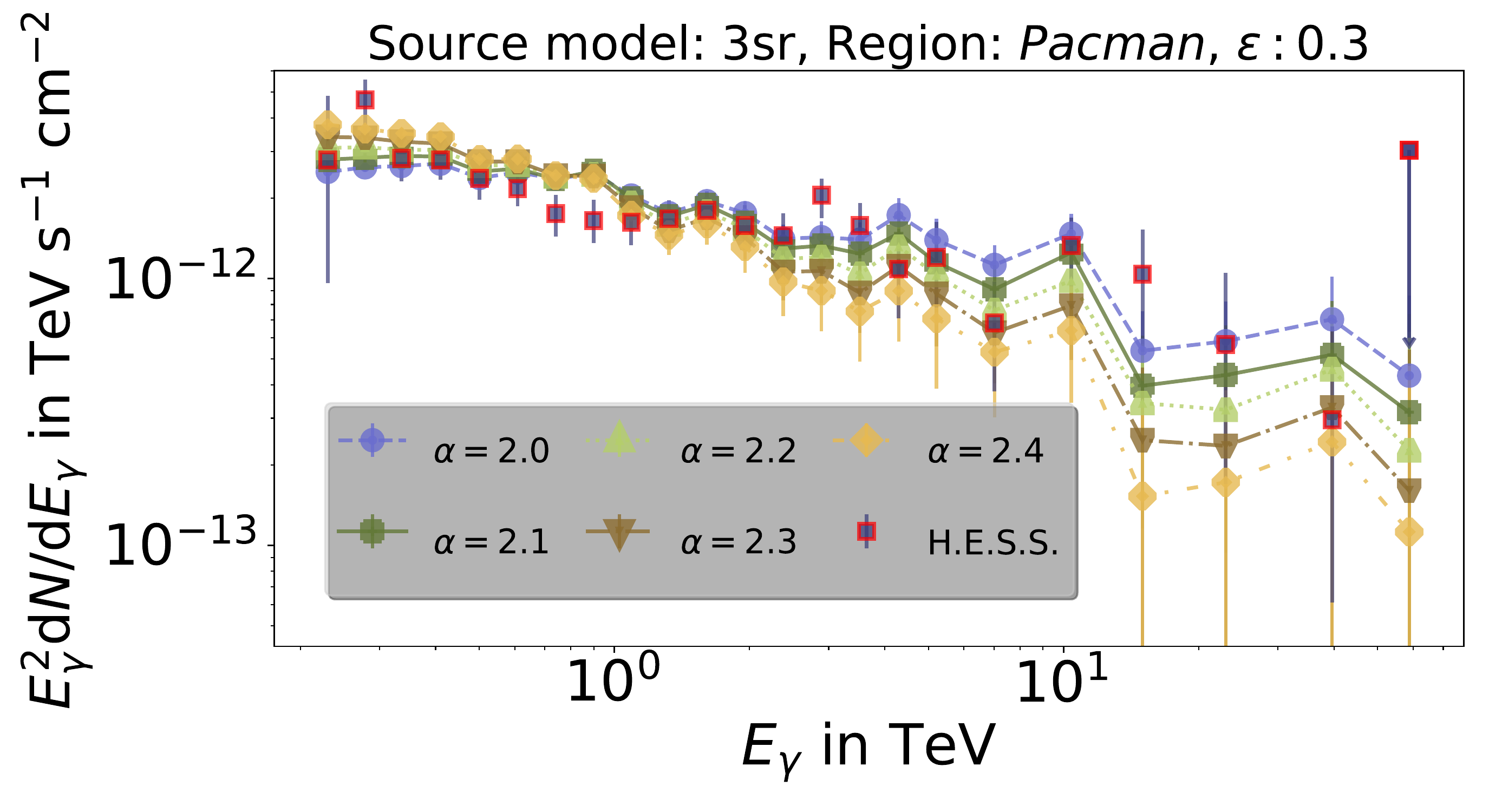}}
	\caption{The differential gamma-ray flux is presented as a function of the energy and for different proton spectral indices and $\epsilon$ values.}
	\label{fig:FluxGamma}
\end{figure*}

In the simulation, we inject a cosmic-ray energy spectrum $Q\propto E^{-2}$ and we reweight this spectrum to different energy behaviors $Q\propto E^{-\alpha}$ with $\alpha\in$[2.0, 2.05, ..., 2.4]. The energy interval binning is adjusted to match H.E.S.S.\ data presented in \cite{AbramowskiNature}. The normalization factor is determined by assuming that the sum of the data bins equals the sum of simulated data bins, i.e. the surface below the data points is the same. The photon energy spectra resulting from the simulations performed here are shown in the four panels of Fig.\ \ref{fig:FluxGamma} - the uppermost panels are for $\epsilon=0.001$ (left) and $\epsilon=0.01$ (right) and the lower ones for $\epsilon=0.1$ (left) and $\epsilon=0.3$ (right). In each panel, the result is shown for five different spectral indices, i.e.\ $\alpha=2.0$ (circle), $\alpha=2.1$ (pentagon), $\alpha=2.2$ (triangle up), $\alpha=2.3$ (triangle down) and $\alpha=2.4$ (diamond). The H.E.S.S.\ data are shown as well (red squares). The best fit to the data is achieved for the combination $\epsilon=0.001$ and $\alpha=2.3$, where we base the goodness of the fit on chi-square tests under the null hypothesis that the observed data are described by the simulation. The results of this test are presented in Table \ref{tab:chisquareSpectral}. An increase of $\epsilon$ seems to decrease the best-fit spectral index. The simplistic assumption by \cite{AbramowskiNature}, i.e. $\alpha=\alpha_{\gamma}-0.1$, requires a proton spectral index of $\sim$ 2.4, which is slightly steeper than what we find. 
 %\clearpage
%\clearpage

Different leptonic and hadronic processes can contribute to the gamma-ray energy spectrum as discussed above. Recently, \cite{Porter2018} predict a significant attenuation of high-energy photons due to their interaction with the ambient photon field. In their calculation, they assume gamma-ray emission from one centralized source. By contrast, this work considers the entire CMZ, and the results of this chapter can be used to test the previous prediction. Moreover, this work takes into account gamma-ray production via inverse Compton scattering by secondary electrons. 

\begin{figure}[ht]
	%	\vspace*{-1cm}
	\centering
	\subfigure{\includegraphics[width=1.0\linewidth]{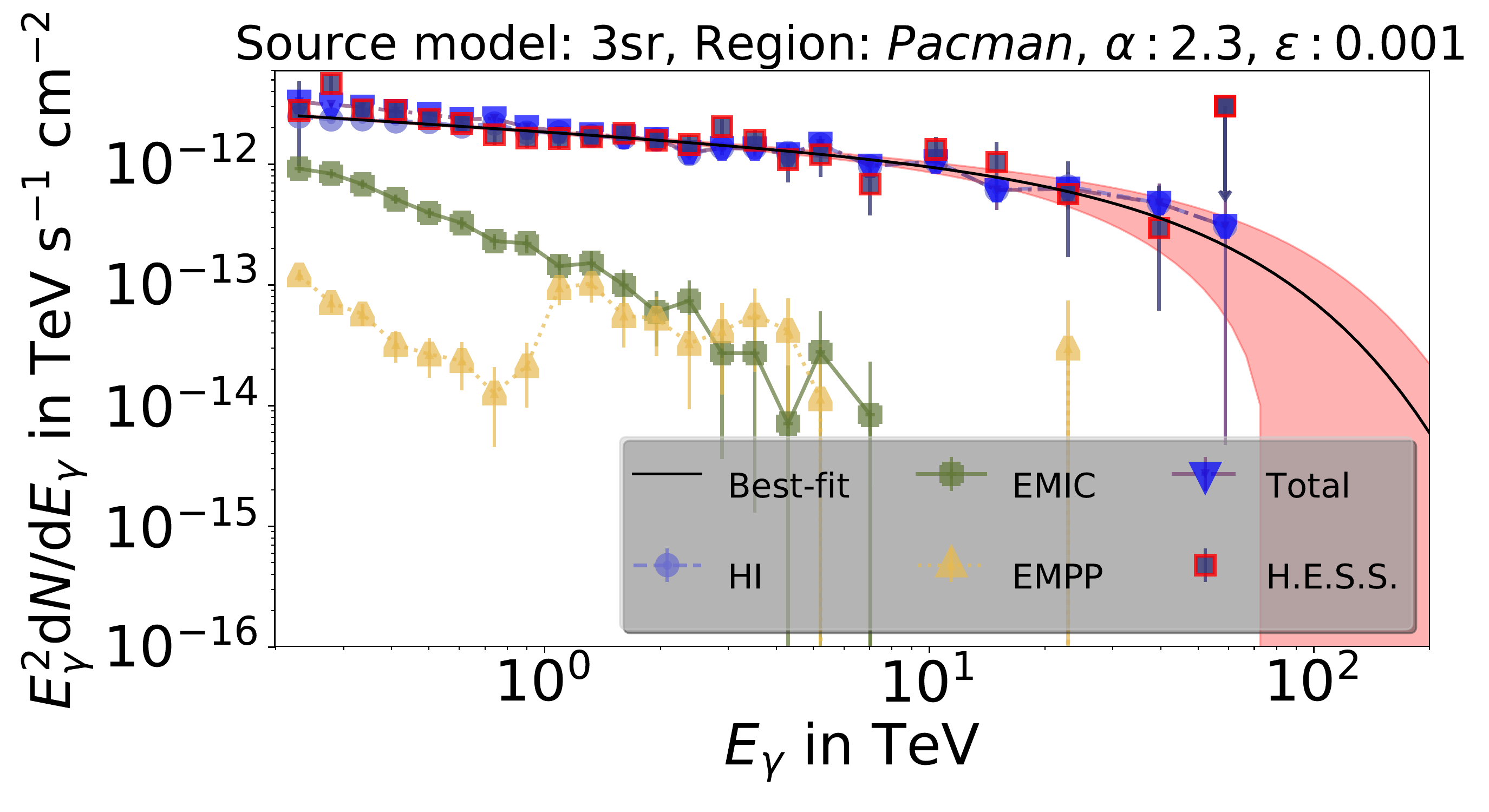}}
	\caption[Best-fit gamma-ray flux obtained from simulations]{The total gamma-ray differential flux with respect to its components is presented. Furthermore, a best fit concerning a power law with cut-off is applied. The red, filled band represents the best-fit $90\,\%$ confidence band.}
	\label{fig:FluxGammaTotal}
\end{figure}

Figure \ref{fig:FluxGammaTotal} displays the total differential  photon flux from the best-fit simulation ($\epsilon=0.001$ with source index $\alpha=2.3$) as well as the contribution from proton-proton interactions ($HI$, circle) and from electromagnetic inverse Compton ($EMIC$, pentagon).

The figure shows that the dominant contribution above $100$~GeV photon energy indeed comes from the hadronic interaction channel. Inverse Compton (IC) scattering has a relatively high contribution at $\sim 100$~GeV yet due to the steepness of the spectrum, the relative contribution of IC decreases rapidly with energy. It is possible that primary electrons still have an impact on the photon spectrum until TeV energies near the center of the Galaxy, but this question is not investigated in this paper, as we do not propagate the electrons here. 

Figure \ref{fig:FluxGammaTotal} even shows the fraction of the originally produced photon flux that is attenuated through gamma-gamma interaction (\textit{EMPP}, triangle up). 

The attenuation reduces the flux by $1\%$ to $10\%$ and occurs mainly at energies below $\sim 7$~TeV. This attenuation is significantly lower compared to the results presented in \cite{Porter2018}. 
A reason could be that the photons in our simulations are produced in a relatively broad region in the CMZ, while \cite{Porter2018} assume a single central source. Thus, the photons in our simulations traverse a smaller column than in \cite{Porter2018}, which leads to a smaller attenuation effect. 

We also perform a fit to both photon, neutrino and electron data from the simulation. We fit the following function for all three spectra:
\begin{equation}
    \frac{\mathrm{d}N_{i}}{\mathrm{d}E}=N_{0}^i\cdot \left( \frac{E}{1\, \mathrm{TeV}} \right)^{-\alpha_{i}} \cdot \exp\left(-\frac{E}{E^i_{\mathrm{cut}}}\right)
    \label{eq:PowerLaw}
\end{equation}
with $i=\gamma,\nu,e$.\\
Using a source emission $\mathrm{d}N/\mathrm{d}E \propto E^{-\alpha}$ with $\alpha = 2.3$ and strong parallel diffusion ($\epsilon = 0.001$) the best-fit adjustment for the gamma-ray flux in Fig.\  \ref{fig:FluxGammaTotal} ($\epsilon=0.001$) delivers $N_{0}^{\gamma}=(1.37+\pm0.03)\cdot10^{-9}$~eV$^{-1}$~cm$^{-2}$s$^{-1}$, $\alpha_{\gamma}=2.20\pm0.02$ and $E^{\gamma}_{\mathrm{cut}}=42.4\pm17.6$~TeV. It should be noted that we do not include a cutoff in our simulations. The cutoff in this spectrum is consistent with being above the last significant simulation data point. A discussion of the cutoff in the data can be found in \cite{AbramowskiNature}. We refrain from discussing this point in detail as we consider the data not to deliver significant information about a cutoff at this point. With CTA data, this will change in the future.

The results for the neutrino--spectra are discussed in Section \ref{neutrinos}.

%================================================================
\subsubsection{Luminosity profile }
%================================================================
In the next step, the simulation results are used for the calculation of the gamma-ray luminosity, which we calculate as
\begin{equation}
    L_{\gamma}=\int_{E_{\min}}^{E_{\max}}\frac{dN_{\gamma}}{dE_{\gamma}}\,E_{\gamma}\,dE_{\gamma}\,.
\end{equation}
Here, we use $E_{\min}=10^{12}$~eV and $E_{\max}=10^{15}$~eV as the boundaries consistent with the H.E.S.S.\ observational range.

\begin{figure}[tb]
	%	\vspace*{-1cm}
	\centering
    \includegraphics[width=1.0\linewidth]{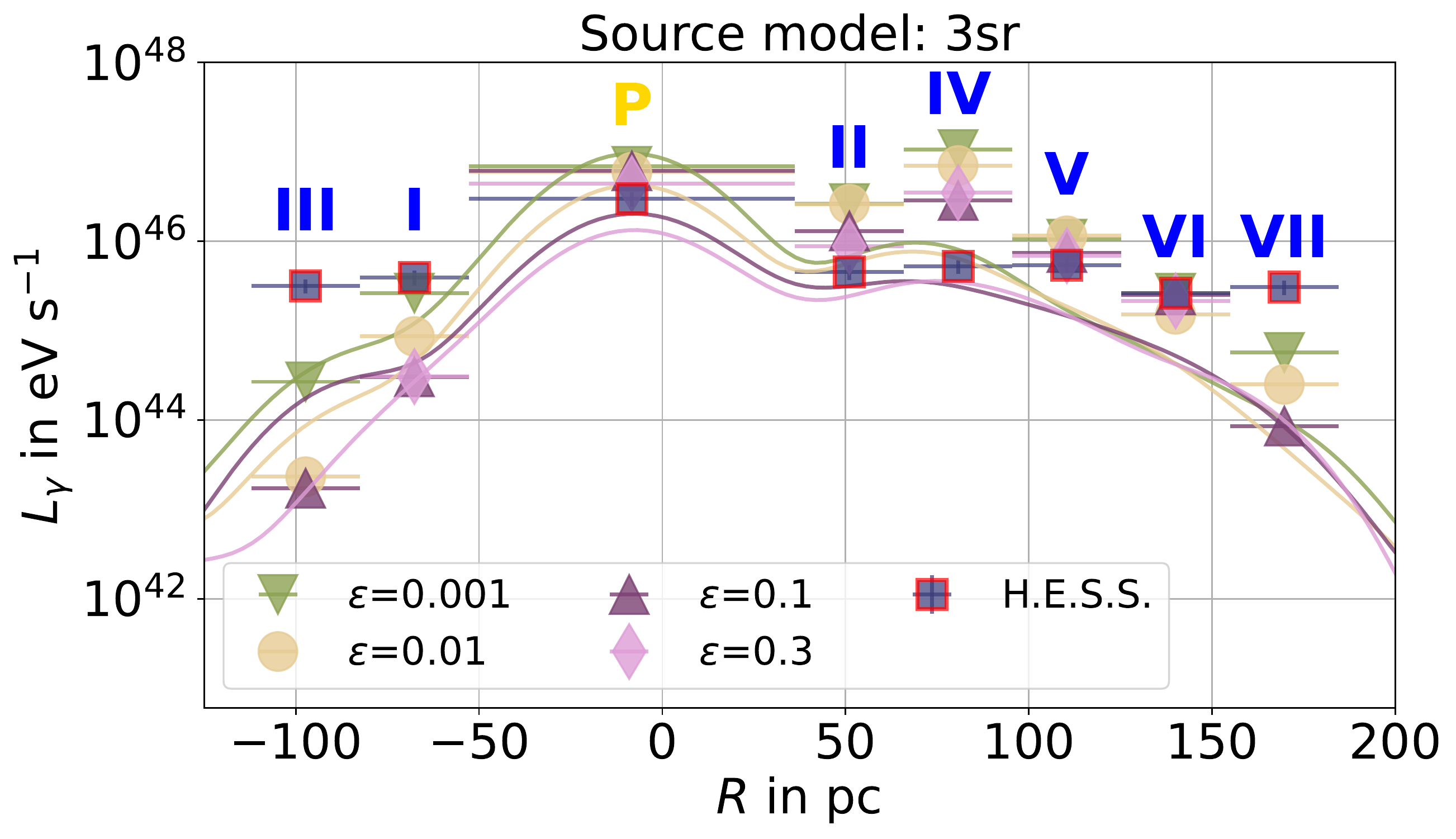}
	\caption{The gamma-ray luminosity from the simulations is presented for the source setup \textbf{[3sr]} and different values of $\epsilon$. The luminosities are obtained for energies $E> 1$~TeV. The H.E.S.S. observation \citep{AbramowskiNature} and the referred region of detection are displayed (compare fig.\ \ref{fig:Regions}).}c
	\label{fig:Luminosity}
\end{figure}

The resulting gamma-ray luminosity as a function of the distance from the point of origin is presented in Fig.\ \ref{fig:Luminosity}. 
The luminosity detected by H.E.S.S.\ from specific extended regions, as described in Figure \ref{fig:Regions}, is presented by blue-filled, red-edged squares in Fig.\ref{fig:Luminosity}. The roman numbers written next to the data points correspond to the labeling of measured areas in Fig.\ \ref{fig:Regions}, also displayed in small in the top-left corner of Fig.\ \ref{fig:Luminosity}. The colored crosses represent our simulation results for the same spatial regions as the H.E.S.S.\ measurements. The different lines represent the continuous luminosity progressions for
$\epsilon = 0.001$ (dashed line), $\epsilon = 0.01$ (solid line), $\epsilon = 0.1$ (dotted line) and $\epsilon = 0.3$ (dash-dotted line), which are integrated over $|b|<0.3^{\circ}$ and smeared to the H.E.S.S. angular resolution of $0.077^{\circ}$. The different gamma-ray luminosities have been normalized using the same normalization factor as found for the differential gamma-ray flux shown in Section \ref{GammaFlux}.\par

In the figure, it can be seen that the central peak can be reproduced well. This result is in contrast to the analytical solution of the problem using a spherically symmetric density presented in \cite{Guenduez18}, where the peak was difficult to reproduce. Comparing the results for different values of $\epsilon$, the smaller values 0.01 and 0.001 describe the data best. In general, the central part is well reproduced, but the mismatch between simulation and data increases toward the outer parts of the CMZ. The largest discrepancy between our model and observational data therefore exist in the outermost data points at the spatial bins III and VII. Such an effect could be an indication for a contribution of the diffuse cosmic-ray flux from sources outside of the GC, something that could be interesting to consider in the future.
%================================================================
\subsection{Gamma-ray count maps}
\begin{figure}[t]
	%	\vspace*{-1cm}
	\centering
	\subfigure{\includegraphics[width=1.0\linewidth]{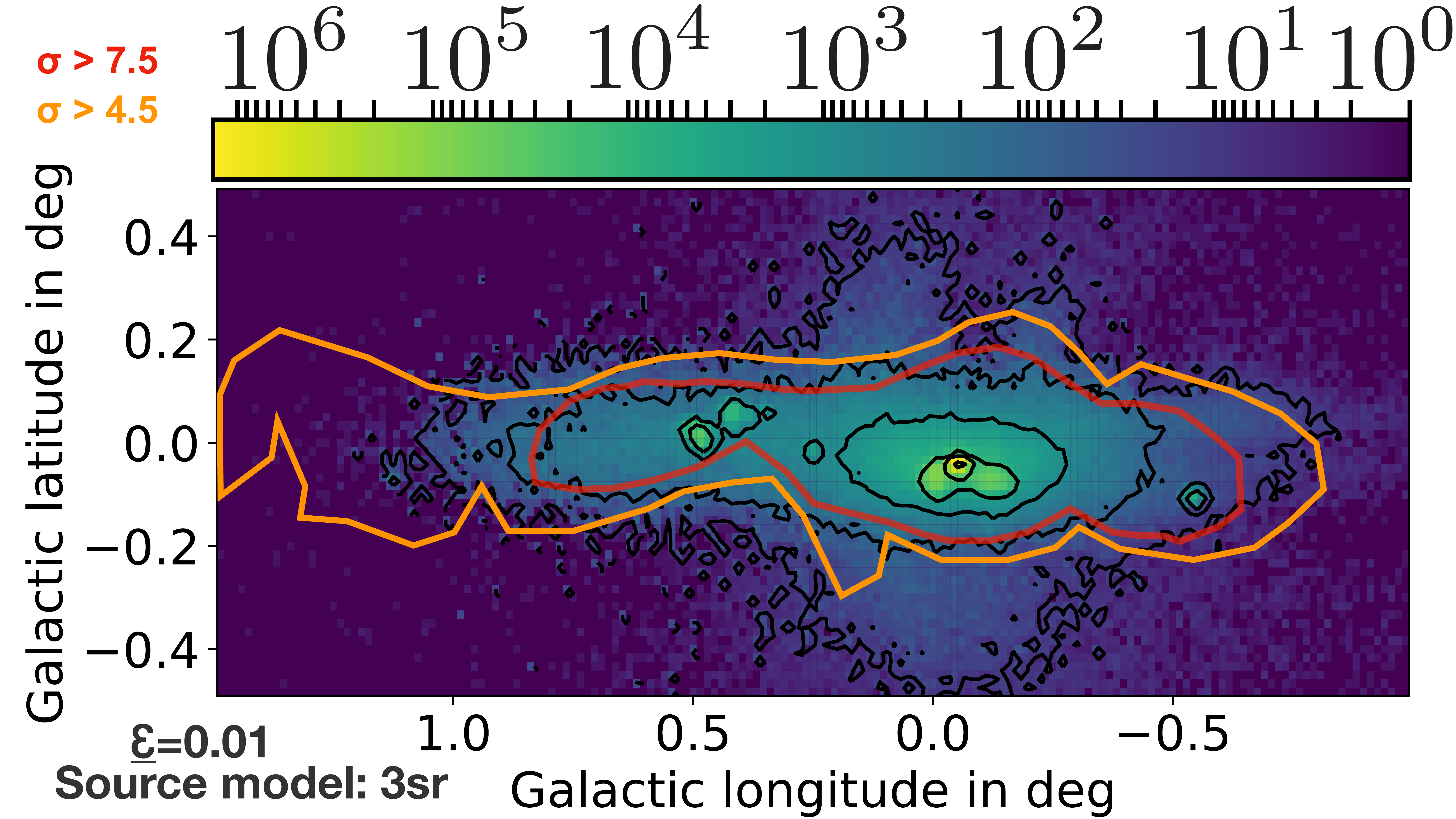}}
	\subfigure{\includegraphics[width=1.0\linewidth]{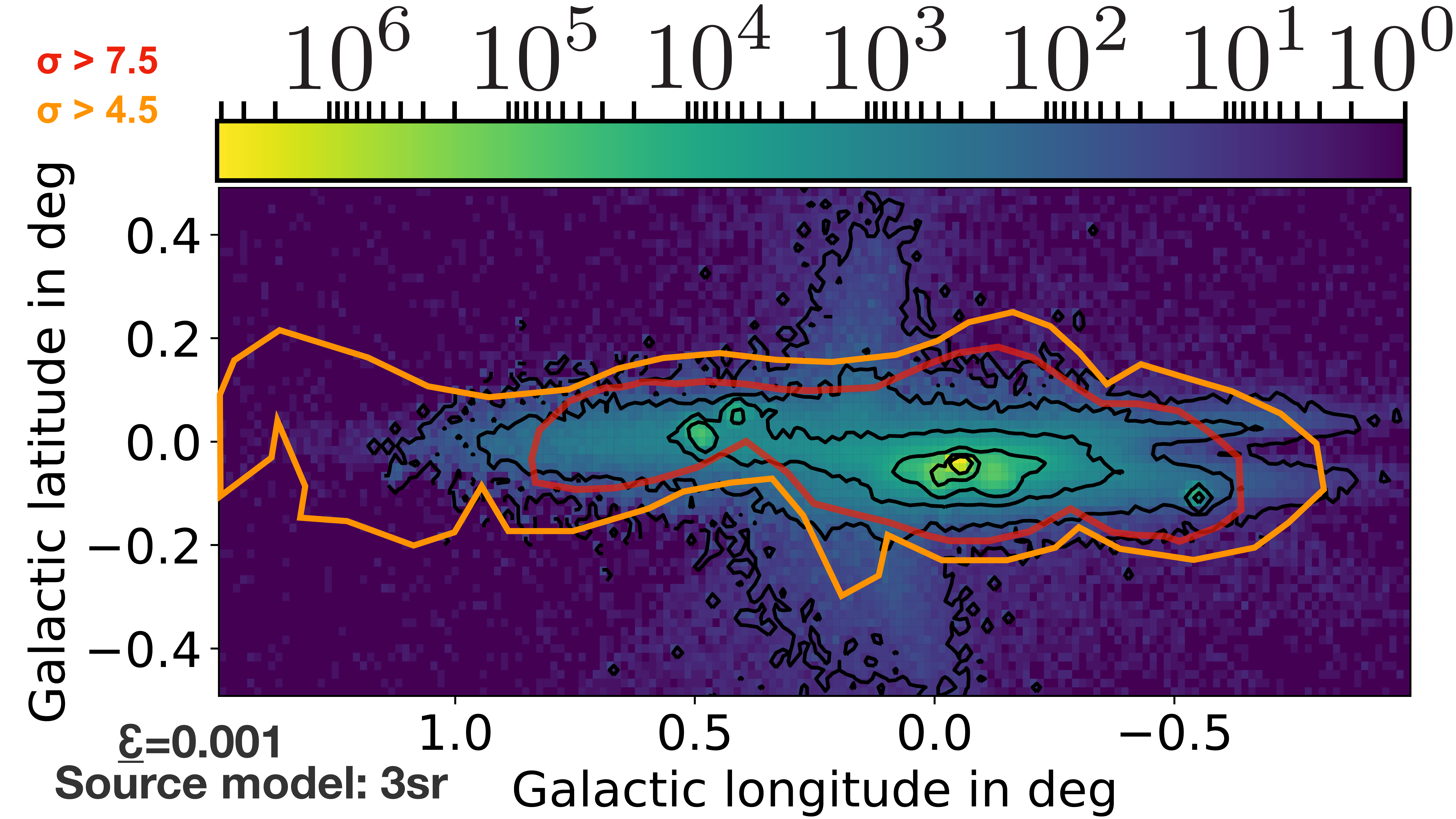}}
	\caption{The gamma-ray count map as predicted here is shown for $\epsilon=0.01$ (upper panel) and $\epsilon=0.001$ (lower panel).  Here, raw data has been used, i.e.\ no smearing is performed.}
	\label{fig:HistoGamma}
\end{figure}

%================================================================
Figure \ref{fig:HistoGamma} shows the count maps of the photons that are produced in the GC via the interaction of the $10^{6}$ induced cosmic-ray protons with the ambient medium for $\epsilon=0.01$ (upper panel) and $\epsilon=0.001$ (lower panel). We find that the center emission is very concentrated and declines strongly toward high Galactic latitudes, basically disappearing at values of $|b|>0.2$. The decline toward larger Galactic longitudes is slower; the signal extends up to $l\sim 1.0^{\circ}$ and disappears at $l<-0.5^{\circ}$ at negative latitudes. 

\begin{figure}[t]
	\centering
	\subfigure{\includegraphics[width=1.0\linewidth]{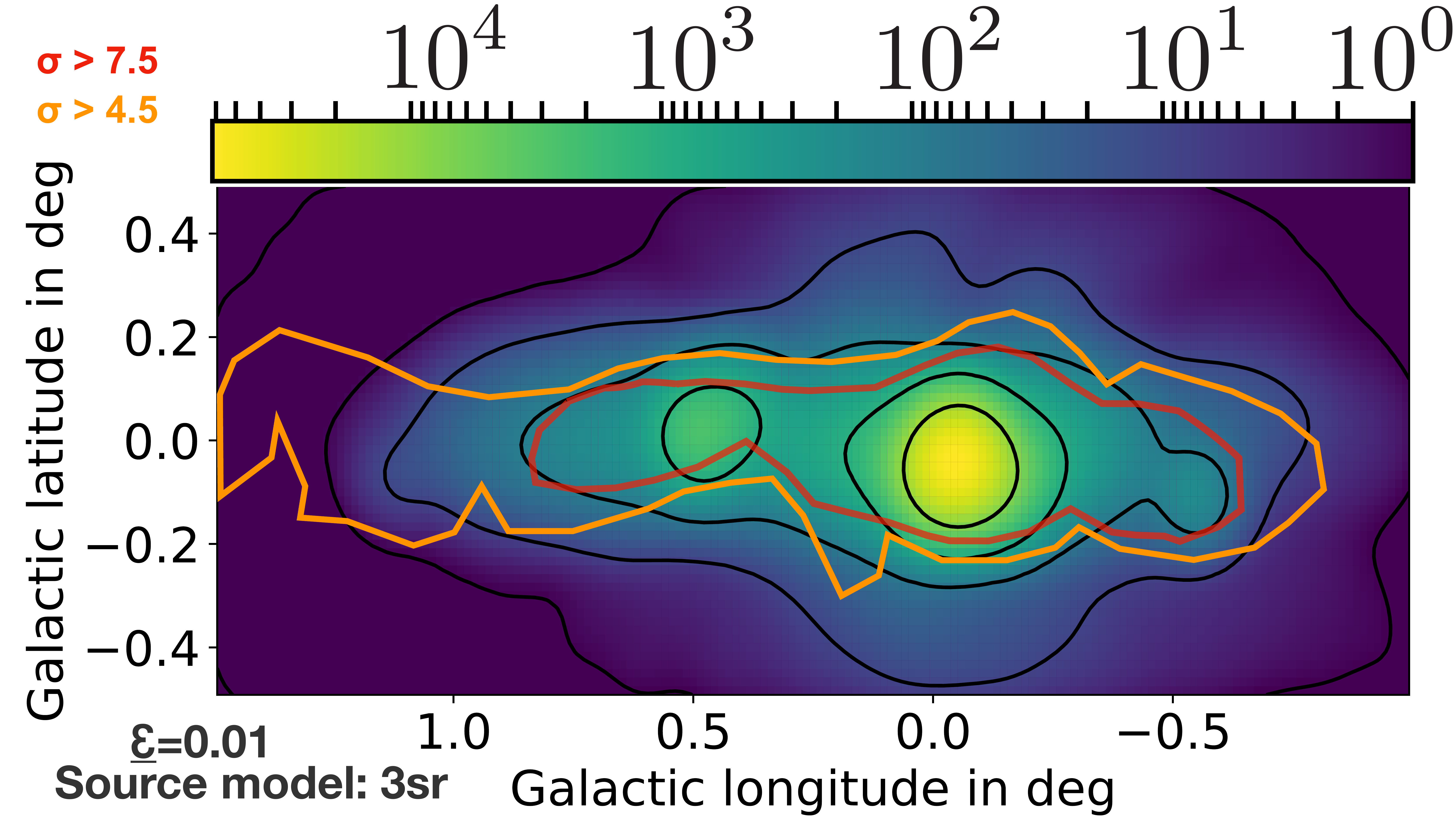}}
	\subfigure{\includegraphics[width=1.0\linewidth]{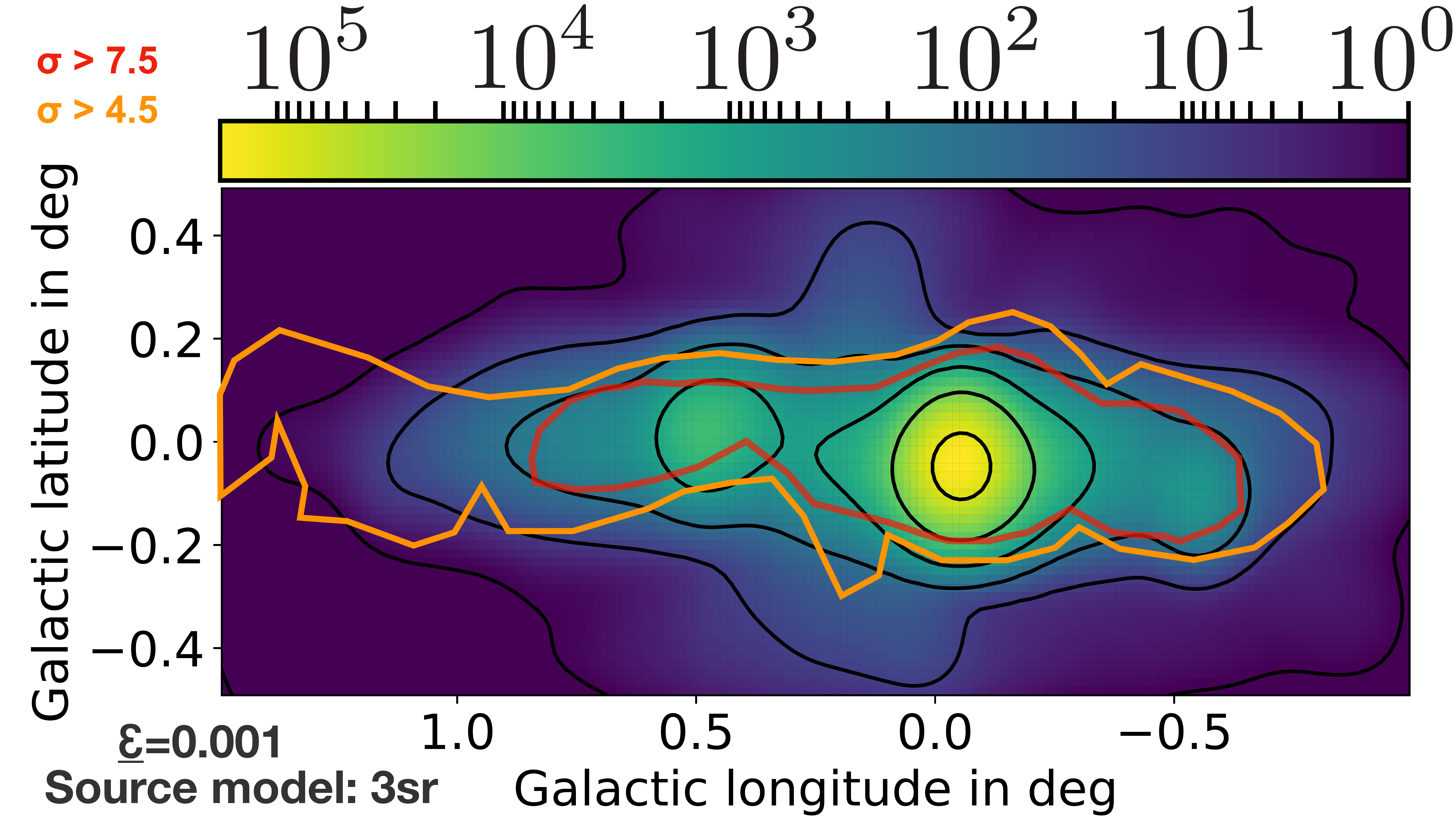}}
	\caption{As Figure \ref{fig:HistoGamma}, but with a smearing corresponding to the H.E.S.S. angular resolution ($0.077^{\circ}$).}
	\label{fig:HistoGammaGaussian}
\end{figure}

In order to properly compare the simulated spatial structure of the gamma-ray map to the emission observed by H.E.S.S., Fig.\ \ref{fig:HistoGammaGaussian} shows our results using \textit{Gaussian smoothing} that corresponds to the H.E.S.S.\ angular resolution of 0.077$^\circ$. The figure also shows the $>4.5\,\sigma$ and $>7.5\,\sigma$ significance levels for the H.E.S.S.\ detection as orange and red lines, respectively. For comparison, the predicted lines of equal gamma-ray counts are added in black. While we do not make a prediction of the normalization itself, it can be seen that the shape of the observed distribution is matched well by the simulation data.  

The region with the highest significance of detection, enclosed by the red line, in general shows quite good agreement with our simulation data.
Even in this representation, the simulation with $\epsilon=0.001$ fits slightly better than $\epsilon=0.01$ or larger epsilon values. As this is consistent with the findings in the previous subsections, we show the following plots for $\epsilon=0.001$ only. The discrepancy is more significant for positive longitudes, consistent with what we found in the results shown in Fig.\ \ref{fig:CountMap}. One reason for this could be existing Molecular Clouds that are not considered in this work due to the lack of data. The existence of such additional MCs has been discussed by \cite{Oka2001}, but it remains difficult to detect a full sample at this point. In particular, the cataloged masses show inconsistencies with other observations, see e.g.\ \cite{CMZMC,MassGalacticCenter2,SgrB2}, which might be due to an overestimation obtained from the usage of virial masses that rather represent an upper limit than an exact measurement. Neglecting the Galactic background diffuse large-scale contribution, as described by \cite{Abramowski2014}, can also partially explain the discrepancy.\par

\begin{figure}[t]
\centering
	\includegraphics[width=1.0\linewidth]{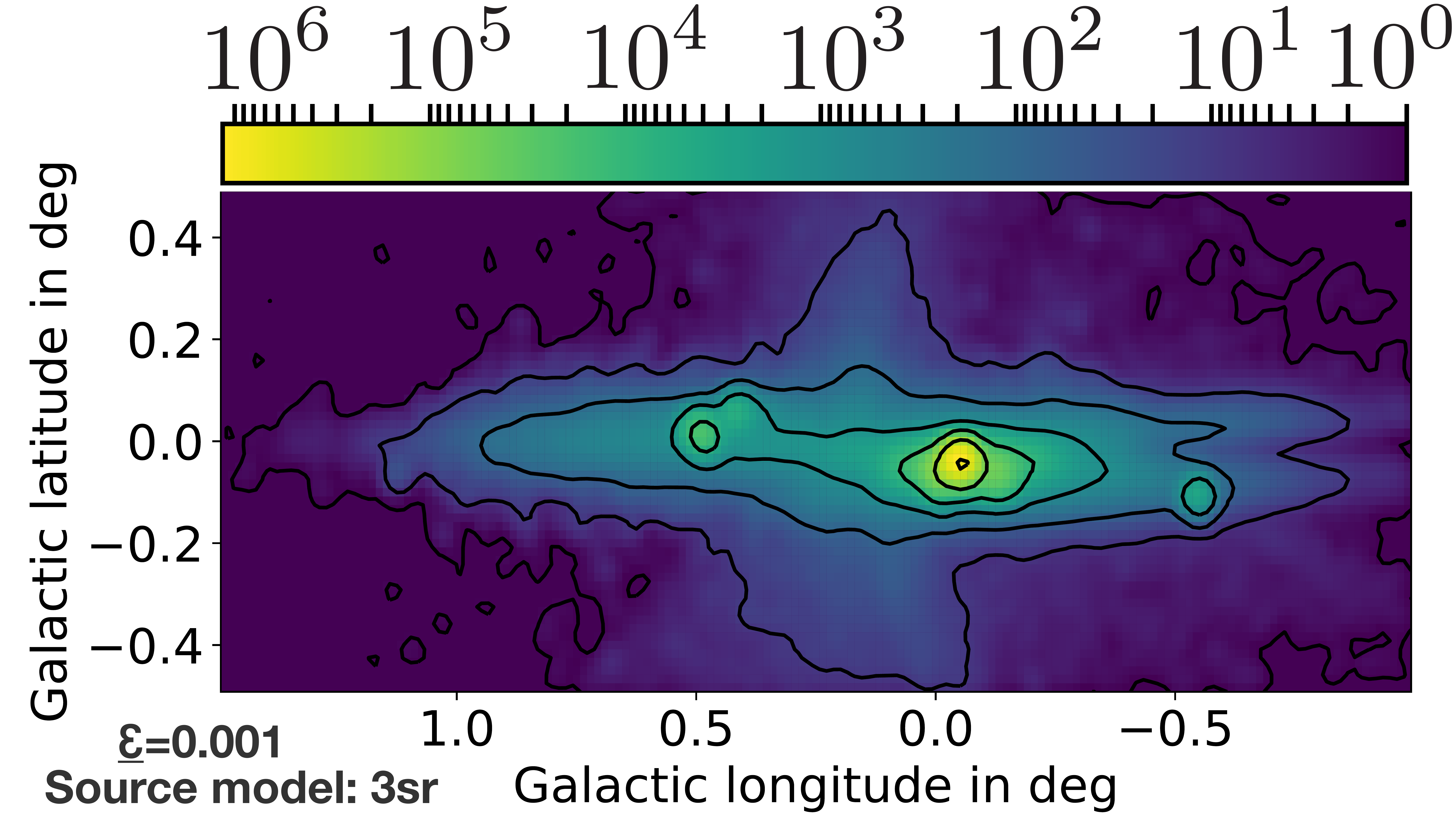}
	\caption{CTA count map (smoothing of $0.03^{\circ})$ using $\epsilon = 0.001$.}
	\label{fig:HistoGammaGaussianCTA}
\end{figure}

Figure \ref{fig:HistoGammaGaussianCTA} shows the gamma-ray map with a smearing of $0.03^{\circ}$ which corresponds to the approximate resolution of CTA south for $>$~TeV energies \citep{cta}. At such a high resolution, the structure of the gas can be identified well. Beyond the central source, there are two specific localized regions that show up at the CTA resolution which are less pronounced for the H.E.S.S.\ data: at a longitude of $b\sim 0.5$, the six dust ridge clouds (A-F) are illuminated. At a longitude of $b\sim -0.5$, the cloud Sgr C shows up. It can be expected that CTA will discover even more molecular clouds, as the analysis of the comparison of our simulation data with the H.E.S.S.\ results in previous subsections indicates that there is still some significant amount of MC gas that could exist, but is not identified at this point.

%================================================================
\subsection{Neutrino count maps and energy spectrum \label{neutrinos}}
%================================================================

\begin{figure}[t]
	\centering
	\subfigure{\includegraphics[width=1.0\linewidth]{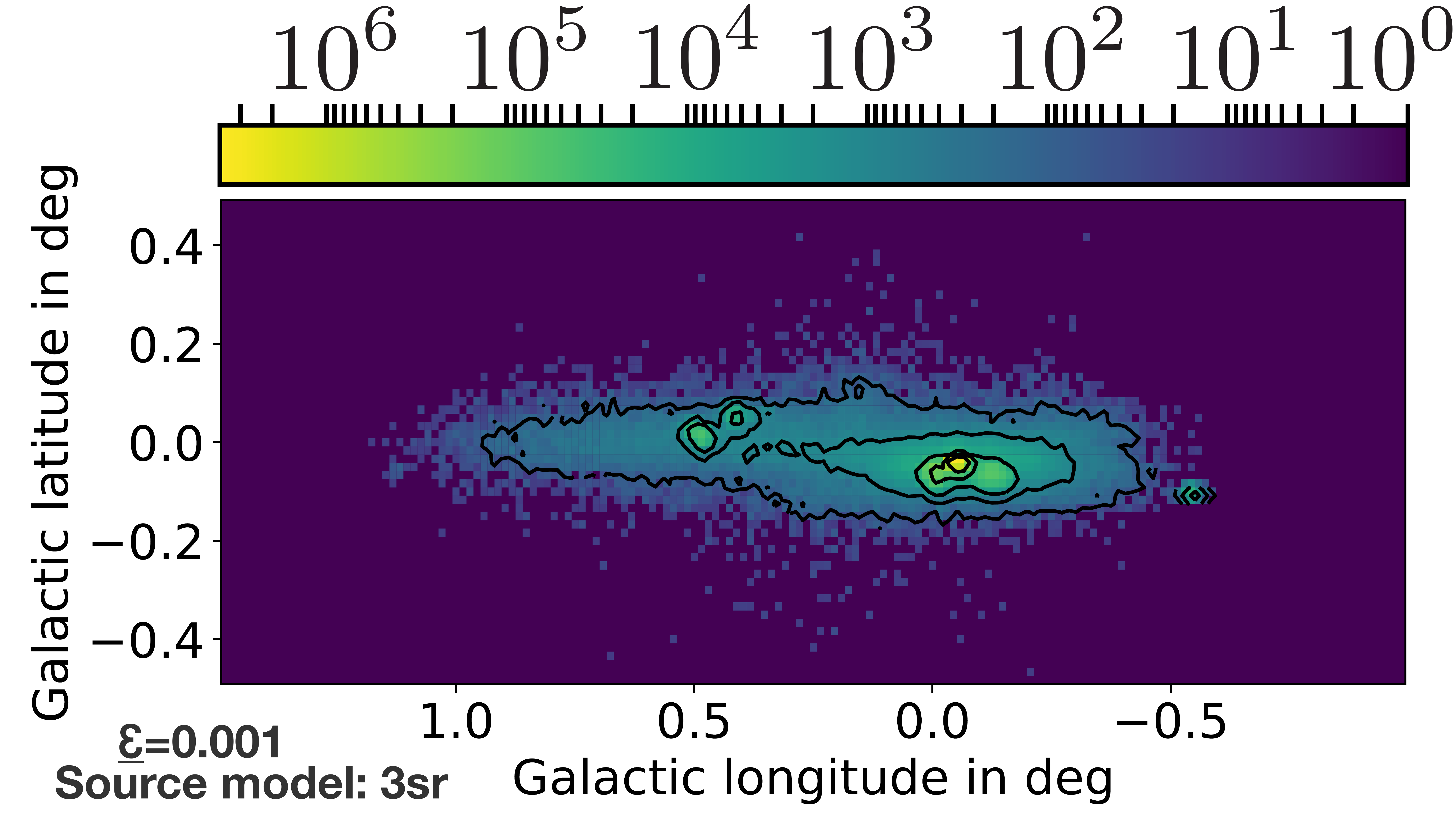}}
	\caption[Neutrino count map in the Galactic Center]{The neutrino count map is shown. The simulation count levels are represented by black lines. Here, raw data has been used, i.e. no smearing is performed.}
	\label{fig:HistoNeutrino}
\end{figure}

The count map of co-produced neutrinos via proton-proton interactions is presented in Fig.\ \ref{fig:HistoNeutrino}. The map shows all neutrino flavors, but considering neutrino oscillations from the ratio at a production of $(\nu_e:\nu_\mu:\nu_\tau)=(1:2:0)$. With neutrino oscillations, the ratio is $(\nu_e;\nu_\mu:\nu_\tau)=(1:1:1)$ so that without considering the normalization, the picture shown does represent the flux of each neutrino flavor individually. This is important, since the best pointing for neutrinos is received for charged current interactions of muon neutrinos in the Antartic Ice instrumented by the IceCube Neutrino Observatory.
 
While the count map is quite similar to the gamma-ray one, it is not exactly the same. There are two reasons for this: first of all, Inverse Compton processes contribute to the gamma-ray map. For the hadronic part, neutrino- and gamma-ray maps should basically be close-to identical at production. Afterwards, gamma rays are absorbed by gamma-gamma interactions, which changes the count map with respect to the one at production. These two effects alter the gamma-ray map with respect to the neutrino one.

\begin{figure}[t]
	\centering
	\subfigure{\includegraphics[width=1.0\linewidth]{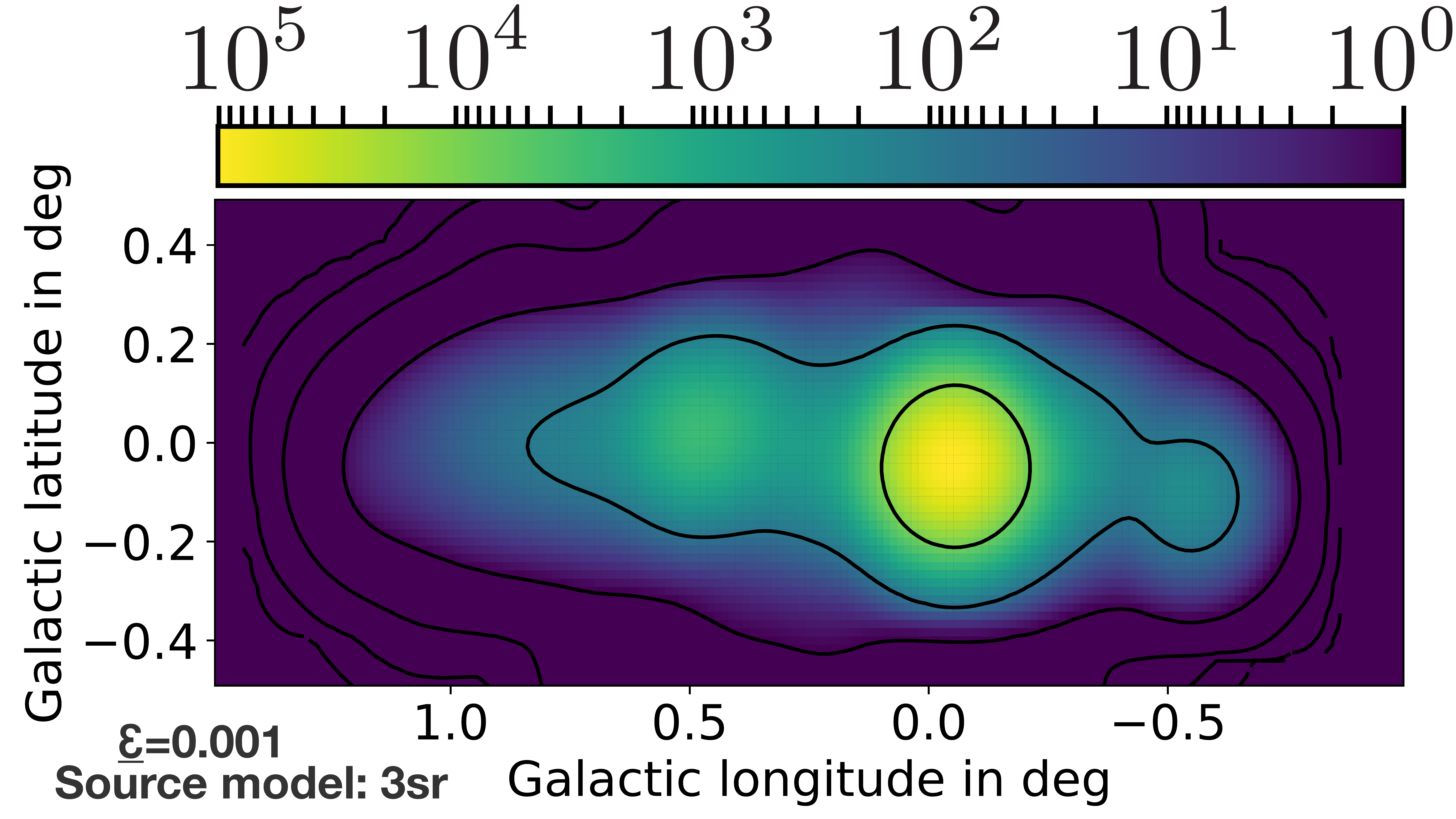}}
	\caption[Smoothed neutrino count map in the Galactic Center]{Figure \ref{fig:HistoNeutrino} is presented considering a smearing corresponding to an angular resolution of 0.1$^{\circ}$.}
	\label{fig:HistoNeutrino2}
\end{figure}

A smeared version with an angular resolution of $0.1^\circ$ is shown in Fig.\ \ref{fig:HistoNeutrino2}. This resolution corresponds to the one of IceCube for \textit{through-going tracks} at $\sim 2$~TeV \citep{IceCube7Years}. Considering only kinematic effects, this resolution is already obtained at $0.1$~TeV, and the resolution is in general increasing with increasing energy, as a result of the enhanced forward scattering at higher energies \citep{learned_mannheim,Julia1}. 

As can be seen from Fig.\  \ref{fig:HistoNeutrino}, it will be difficult to see sub-structures, even if the number of detected neutrinos would be enough to have a spatial resolution. So, IceCube right now will have difficulties in resolving these structure, with future observatories and detections at $\sim 100$~TeV$ -- $~PeV, the resolution becomes better and the substructures might emerge, if the event rates are large enough (see further discussion below).%As we discuss in the upcoming sub-section, the sensitivity of current and next generation neutrino telescopes is not sufficient for a detailed spatial resolution, so that such a neutrino map is not expected to be obtained in data in the near future.

\begin{figure}[ht]
	\centering
	\includegraphics[width=0.9\linewidth]{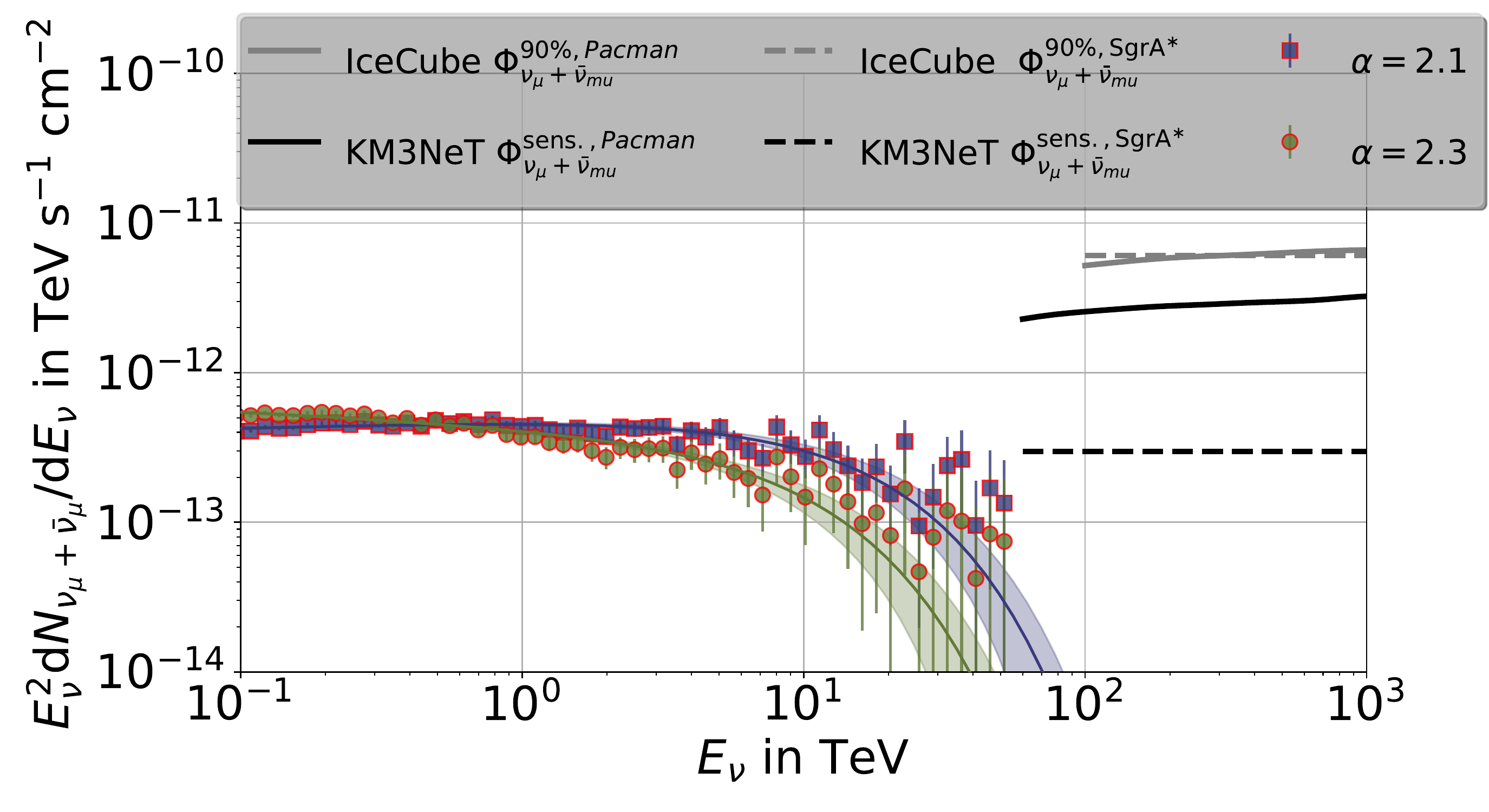}
	\caption{The prediction for the neutrino flux from the Galactic Center region for primary spectra of $E^{-2.1}$ (green squares) and $E^{-2.3}$ (blue circles). Power-law fits with cutoff are performed and shown as solid lines. The grey lines show IceCube 90\% confidence limits for \SgrA (dashed) and the Pacman region (solid) for 10 years of data taking. The black lines show the KM3NeT sensitivity for the 90\% confidence level in 6 years of data taking, for \SgrA\ (dashed) and the Pacman region (solid).}
	\label{fig:FluxNeutrino}
\end{figure}

The neutrino energy spectrum is calculated in the same way as the gamma-ray energy spectrum (see Section \ref{GammaFlux}). In Fig.\ \ref{fig:FluxNeutrino}, the simulated energy spectrum of muon neutrinos and anti-muon neutrinos is shown, derived from the flux of all neutrinos by considering neutrino oscillations from a flavor ratio at production of $\left.(\nu_{e}:\nu_{\mu}:\nu_{\tau})\right|_{\rm source}=(1:2:0)$ to the one at Earth of $\left.(\nu_{e}:\nu_{\mu}:\nu_{\tau})\right|_{\oplus}=(1:1:1)$. We show the simulation results for primary spectra of $\alpha=2.1$ (squares) and $\alpha=2.3$ (circles).
The power-law fit (Equ.\ \ref{eq:PowerLaw}) for the neutrino flux induced by a source population with a primary energy spectrum of $E^{-2.1}$ and $E^{-2.3}$ yields the parameters given in table \ref{tab:NeutrinoFit}.

\renewcommand{\arraystretch}{1.2}
\begin{table*}[htb]
    \centering
    \caption{Power-law fit (Equ.\ \ref{eq:PowerLaw}) results for the neutrino flux shown in fig.\ \ref{fig:FluxNeutrino}.}
    \label{tab:NeutrinoFit}
    \begin{tabular}{||c||c|c||}
    \hline
        $\alpha_\mathrm{source}$ & $2.1$ & $2.3$  \\ \hline \hline \rule{0pt}{1.2\normalbaselineskip}
        $N_0^\nu$ $[10^{-15} \ \mathrm{TeV^{-1} cm^{-2} s^{-1}}$] & $478 \pm 9$ & $441 \pm 12$ \\ \hline
        $\alpha$ & $1.951 \pm 0.011$ & $2.095 \pm 0.014$ \\ \hline
        $E_\mathrm{cut} \ [\mathrm{TeV}]$ & $17.3 \pm 2.4$ & $11.3 \pm 2.1$ \\ \hline \hline
    \end{tabular}
\end{table*}

As for the gamma-ray spectrum, the cutoff-energy is not an effect of a cutoff in the simulation, but it is located at energies that are higher than the simulated, significant data points. It is basically a lower-limit for a cutoff in the spectrum.

An IceCube $90\,\%$ C.L. upper flux limit for the position of \SgrA\ ($\sin(\delta_{SgrA*}=-29.01^{\circ})=-48.5$) is given in \cite{icecube_10yrps} for a spectral index of $\alpha=$2.0 (dashed, grey line in Fig.\ \ref{fig:FluxNeutrino}). The lower energy threshold of the analysis is relatively high, as the GC region is located at a declination quite far above the horizon at the South Pole, i.e.\ $\delta=-30^{\circ}$. This implies that there is a large background of atmospheric muons, which is reduced by applying the high-energy cut. A different method is to only consider events that start inside of the detector. This approach was first developed for high energies ("high energy starting events", HESE) and lead to the first detection of the diffuse neutrino flux \citep{IceCubeScience2013}. The extension of this method to lower energies followed shortly after \citep{IceCube_MESE} and provides another channel of detection of the Galactic Center region.

The expected neutrino flux from the Galactic Center emission can be converted into a number of neutrinos to be detected in different analyses with IceCube based on the effective areas. Here we consider only the muon-neutrinos and their effective area, as the direction uncertainty for the other flavors is to large. 
For the HESE sample it is published for the declination bands $[0^\circ;-30^{\circ}]$ and $[-30^{\circ} ; -90^{\circ}]$ concerning the point source search \citep{IceCube_PointSource}.  As the Galactic Center lies at the boundary of these two declination bands $\delta \sim -29^{\circ}$, we use these two effective areas to calculate the rate of neutrinos according to

\begin{equation}
N_{\nu} =\Delta t_{\rm obs} \int \frac{\mathrm{d}N}{\mathrm{d}E_\nu}\,A_{\rm eff}(E)\,\mathrm{d}E_\nu\,,
\end{equation}
with $\Delta t_{\rm obs}$ as the observation time. Based on this estimate, we provide the range of neutrinos. We receive a total number of neutrinos between $2\cdot 10^{-4}-0.08$ for the full 10 years.
For the Medium Energy Starting Events (MESE) effective area, published in \cite{IceCube_MESE} as supplemental material, we use the zenith angle bin $-0.2 \leq \cos(\theta) \leq 0$, where $\theta = 90^\circ - \delta$ is the zenith angle.
Assuming an observation time of 10 years we receive $\sim 0.03$ events\footnote{Note that the published MESE analysis is based on 4 years of data, so the expected number of events for the published analysis is even smaller, $\sim 0.012$}. Thus, the sources in the Galactic Center itself are not observable with IceCube at this point.. It should be noted that we do not include the diffuse sea of background cosmic rays that propagate into the Galactic Center from outer regions of the Galaxy, so our numbers are rather considered to be a lower limit.

KM3NeT will have a better detection potential due to its location at the opposite side of Earth compared to IceCube. The sensitivity corresponding to the median $90\,\%$ confidence level (C.L.) upper-flux limit for an unbroken power-law distribution proportional to $E^{-2}$ using 6 years of data results in $\Phi^{90\%}_{\nu_{\mu}+\overline{\nu}_{\mu}}=0.297$ eV cm$^{-2}$ s$^{-1}$ \citep{Aiello2019}. Considering the extended source analysis of KM3NeT by \cite{Ambrogi2018}, the minimum detectable flux is reached at $\sim$ 60 TeV. For lower energies, the flux limit increases rapidly. For energies $>$ 60 TeV, the median angular resolution improves from $\sim$ 0.13$^{\circ}$ up to 0.07$^{\circ}$ at $10^5$ TeV. Because the angular resolution is smaller than the extent of the \textit{Pacman} region, the upper limit analyzed for the point source \SgrA\ can be extrapolated to a more extended region by considering the median angular resolution following
\begin{equation}
    \Phi^{\mathrm{Pacman}}_{\nu_{\mu}+\overline{\nu}_{\mu}}\approx\Phi_{\nu_{\mu}+\overline{\nu}_{\mu}}\cdot \frac{\Delta \delta_{\mathrm{Pacman}}}{\delta_{\mathrm{MA}}(E)}\, .
    \label{eq:UpperLimit}
\end{equation}
Here, $\Delta \delta_{\mathrm{Pacman}}$ denotes the extent of the region of interest in degrees and $\delta_{\mathrm{MA}}(E)$ the median angular resolution as a function of the energy. The median angular resolution and KM3NeT using 6 years of data and of IceCube using 7 years of data is presented in \cite{KMNeT} and \cite{IceCube7Years}, respectively. Figure \ref{fig:FluxNeutrino} shows the $90\,\%$ C.L. upper-flux limit of IceCube and the corresponding sensitivity of KM3NeT. The limit calculation follows \cite{Ambrogi2018}, who calculate the sensitivites to extended sources, where the Pacman region has an extension of $\sim 0.5^{\circ}$.

%%%%%%%%%%%%%%%%%%%%%%%%%%%%%%%%%%%%%%%%%%%%%%%%%%%%%%%%%
\section{Summary and Conclusions\label{discussion:sec}}
%%%%%%%%%%%%%%%%%%%%%%%%%%%%%%%%%%%%%%%%%%%%%%%%%%%%%%%%%
In this paper, we apply a 3D propagation model with anisotropic diffusion  in the Galactic Center region, for the first time using a realistic representation of the three-dimensional gas distribution and large-scale magnetic field in this region of the Galaxy. We perform our simulations within the diffusion module of the \textit{CRPropa} framework, using a new implementation of a generalized parametrization of the photon fields, this way being able to include interactions of the high-energy gamma-rays with the photon field coming from the populations of stars in the Galctic Center region. We consider Inverse Compton scattering and synchrotron radiation for the secondary electrons and most importantly, proton-proton interactions for the hadronic production of gamma rays and neutrinos. In doing so, we test five different source distributions for four different ratios of perpendicular to parallel diffusion. We normalize and compare our results to the spatial distribution and energy spectrum of the H.E.S.S.\ detection. We draw the following conclusions
\begin{enumerate}
    \item
    The broad distribution of the H.E.S.S.\ signal is well-represented by a three-source model, including a central source like \SgrA\, a source at the position of SNR G0.9+0.1 and HESS J1746-285. Other source scenarios that we test neither match the two-dimensional distribution nor the one-dimensional projections in longitude and latitude as good. In particular, the distribution of pulsars in the GC does not provide a good match to the data, neither does the sole consideration of \SgrA.

    \item
    The best results are achieved for small perpendicular diffusion values, i.e.\ for a ratio of perpendicular to parallel diffusion of $\epsilon=0.001$.

    \item
    Some small-scale features cannot be reproduced with the model, which points towards the existence of molecular clouds that are not well identified in the data yet and, thus, not part of the three-dimensional mass distribution used.

    \item
    At the same time, the general features are explained well by the source and target distribution, including clouds like Sgr~B2 and the six dust ridge clouds (A-F). While the latter contribute to the diffuse picture in the current measurements, they can be detected as individual gamma-ray emitting regions in the future with the better spatial resolution of CTA.

    \item
    We predict the distribution of neutrinos and note that the resolution of IceCube is not enough to resolve the individual features at this point, together with the fact that the rate of neutrinos is simply too small. For future observatories like KM3NeT and IceCube-Gen2, a detection can be possible, depending on how low in energy threshold the observatories will reach and to what energies particles are accelerated. CTA, with its expected sensitivity at $\sim 100$~TeV photon energy, will be able to provide further insights to the exact flux at the highest energies.
\end{enumerate}

\begin{acknowledgements}
	We would like to thank Isabelle Grenier for valuable discussions on the importance of the photon field for gamma-gamma absorption in the Galactic Center region.  We acknowledge the support from the DFG via the Collaborative Research Center SFB1491 \textit{Cosmic Interacting Matters - From Source to Signal} (JBT, P-SB, JD, HF, AF). We further acknowledge the support from Studienstiftung des Deutschen Volkes (MH) and Rosa-Luxemburg Stiftung (EMZ).
\end{acknowledgements}

%\begin{enumerate}
%	\item
%